\documentclass{aa}
\usepackage{txfonts}
\usepackage{graphicx}
\usepackage{booktabs}
\usepackage{amsmath}
\usepackage{colortbl}
\usepackage{subfig}
\usepackage{rotating}
\usepackage{amssymb}
\usepackage{latexsym}
\usepackage{ifthen}
\usepackage{enumitem}
\begin{document}

\title{Galaxy-wide outflows in z$\sim$1.5 luminous obscured QSOs revealed through NIR slit-resolved spectroscopy}

\author{M. Perna
		\inst{\ref{i1},\ref{i2}}\thanks{E-mail: michele.perna4@unibo.it}
		\and 
	M. Brusa
		\inst{\ref{i1},\ref{i2},\ref{i3}}\thanks{E-mail: marcella.brusa3@unibo.it} 
		\and
	G. Cresci
		\inst{\ref{i4}} 
	\and
	A. Comastri
		\inst{\ref{i2}}
	\and
	G. Lanzuisi
		\inst{\ref{i1},\ref{i2}}	
		\and
	E. Lusso
		\inst{\ref{i4}} 
		\and
	A. Marconi
		\inst{\ref{i5}} 
		\and
	M. Salvato
		\inst{\ref{i3},\ref{iM}}
		\and
	G. Zamorani
		\inst{\ref{i2}}
	\and 
	A. Bongiorno,
		\inst{\ref{i6}}
	\and
	V. Mainieri
		\inst{\ref{i7}} 
	\and
	R. Maiolino
		\inst{\ref{i8}}
	\and 
	M. Mignoli
		\inst{\ref{i2}}
}
\institute{Dipartimento di Fisica e Astronomia, Universit\`a di Bologna, viale Berti Pichat 6/2, 40127 Bologna, Italy\label{i1}
	\and
	INAF - Osservatorio Astronomico di Bologna, via Ranzani 1, 40127 Bologna, Italy\label{i2}
	\and
	Max Planck Institut fur Extraterrestrische Physik, Giessenbachstrasse 1, 85748 Garching bei M\"unchen, Germany\label{i3}
	\and
	INAF - Osservatorio Astrofisico di Arcetri, Largo Enrico Fermi 5, 50125 Firenze, Italy\label{i4}
	\and
	Dipartimento di Fisica e Astronomia, Universit\`a degli Studi di Firenze, via G.Sansone 1, 50019, Sesto Fiorentino (Firenze), Italy\label{i5}
	\and
        Excellence Cluster Universe, Boltzmannstrasse 2, D-85748 Garching bei M\"unchen, Germany\label{iM}
	\and
	European Southern Observatory, Karl-Schwarzschild-str. 2,  85748 Garching bei M\"unchen, Germany\label{i7}
	\and
	Cavendish Laboratory, University of Cambridge, 19 J. J. Thomson Ave., Cambridge CB3 0HE, UK\label{i8}
        \and
	INAF - Osservatorio Astronomico di Roma, via Frascati 33,   00044 Monte Porzio Catone (RM), Italy\label{i6}
}

\date{Accepted 1988 December 15. Received 1988 December 14; in original form 1988 October 11}

%\pagerange{\pageref{firstpage}--\pageref{lastpage}} \pubyear{2002}

\label{firstpage}

\abstract
% context
{}
% aim
{The co-evolution of galaxies and super massive black holes (SMBHs) requires that
some sort of feedback mechanism is operating during the active galactic nuclei (AGN) phases. AGN driven winds are the most likely candidates for such feedback mechanism, but direct observational evidence of their existence and of their effects on the host galaxies are still scarce and their physical origin is still hotly debated. }
% Method
{X-shooter observations of a sample of X--ray selected, obscured quasars at z$\sim$1.5, selected on the basis of their observed red colors and X--ray-to-optical flux ratio, have shown the presence of outflowing ionized gas identified by broad [OIII] emission lines in 6 out of 8 objects, confirming the efficiency of the selection criteria. Here we present slit-resolved spectroscopy for the two brightest sources, XID2028 and XID5321, to study the complex emission and absorption line kinematics.}
% results
%While the blueshifted absorption lines of XID2028 confirm the presence of an outflow close to the line of sight, the redshifted absorption lines of XID5321 could indicate the existence of an object approaching the AGN, and/or a double nucleus in the galaxy.
{We detect outflow extended out to $\sim$ 10 kpc from the central black hole (BH), both as blueshifted and redshifted emission. Interestingly, we also detect kpc scale outflows in the [OII] emission lines and in the neutral gas component, traced by the sodium D and magnesium absorption lines, confirming that a substantial amount of the outflowing mass is in the form of neutral gas.}
% conclusions
{The measured gas velocities and the outflow kinetic powers, inferred under reasonable assumptions on the geometry and physical properties of these two systems, favor an AGN origin for the observed winds.}
%%%%

\maketitle
\section[Introduction]{Introduction}

Many of the most successful models of galaxy formation \citep[e.g.,][]{Hopkins2006a,King2005%,Lapi2014
} require energetic outflows, driven by active galactic nuclei (AGN), and extending over galaxy scales (i.e., $\sim$ 1-10 kpc) to reproduce the scaling relations observed in the local universe between host galaxies and black hole properties \citep[e.g.,][]{Magorrian1998,Marconi2003,Ferrarese2005,Kormendy2013}. % A similar physical mechanism is also invoked to explain the physical properties of the most massive galaxies, and in particular the bright tail of the stellar mass function (\citealt{Croton2006,Bower2006}).

According to the most popular AGN-galaxies co-evolutionary models \citep[e.g.,][]{Hopkins2006a,Menci2008%,Lapi2014
}, the first phase of the quasi-stellar objects (QSOs) life is associated to rapid supermassive black hole (SMBH) growth and efficient star formation (SF) in a dust-enshrouded, dense environment: in this phase AGN are dust and gas obscured. This is followed by the ``blow-out'' or
``feedback'' phase during which the SMBH releases radiative and kinetic energy in the form of powerful, outflowing wind, before becoming a ``normal'' unobscured QSO once the obscuring material has been cleared out (see, e.g., \citealt{Hopkins2008}). In this framework, AGN feedback should be revealed through the presence of outflowing material driven by energetic winds (see, e.g., \citealt{Fabian2012,Zubovas2012}). However, these outflows could also be powered by SF activity \citep[e.g.,][]{Heckman1990,Veilleux2005} and it is not fully understood the relative dominance of the two different mechanisms involved. Regardless of their nature, during the peak of star formation (z$\sim$ 1.5-3) outflows are believed to play a pivotal role in shaping galaxies, because they regulate both SF and black hole growth.

%While powerful outflows sustained by kinetic energy ejected in the hosts of luminous radio-galaxies by relativistic and collimated jets have been commonly detected out to z $\sim$ 4 (see e.g., Nesvabda et al. 2008, 2011), 
The SF- and AGN-induced outflows have recently been found at low-z,
%Spatially resolved IR and mm spectroscopic studies showed the first evidences for the existence of radiatively driven winds at both low-z, 
involving outflowing mass in the form of molecular gas \citep[e.g.,][]{Feruglio2010,Sturm2011,Cicone2012,Cicone2014}, neutral \citep[e.g.,][]{Rupke2011,Rupke2013}, and ionized gas \citep[e.g.,][]{Westmoquette2012,Rupke2013,Zaurin2013,Harrison2012}, as well as at high-z, in QSOs  \citep[][]{Maiolino2012,Harrison2012,CanoDiaz2012} 
and in massive SF galaxies \citep[][]{Schreiber2014,Genzel2014}. The Sloan digital sky survey (SDSS) enabled the selection of candidate sources hosings outflows at z$<0.5$ %a crucial role in understanding  
%it was possible to understand these phenomena in the ``local'' Universe (z$<0.5$) %has been played 
%by the study of selecting sources 
on the basis of well-defined spectral signatures, such as broad FWHM in the [OIII]$\lambda\lambda$4959,5007 lines and/or double peaked profiles that can be ascribed to complex underlying kinematics \citep[e.g.,][]{Liu2010,Fu2012}. %(see e.g., Liu et al. 2010, Fu et al. 2012, Villar-Martin et al. 2011) 
In most cases, the targets have been followed-up with integral field unit (IFU) instruments to assess the existence and the real extension of the outflows \citep[e.g.,][]{Greene2012,Harrison2014}. In recent years, this detailed study has also been extended to higher redshifts (z$>1$) with spatially resolved spectroscopy with high resolution spectrograph and/or near infrared IFU instruments \citep[see, e.g.,][]{Harrison2012}. %,Carniani2014}. 
However, most of the results at high-z are still based on the observations of very luminous, unobscured QSOs, in which the blow-out phase is expected to be near its end, or in galaxies with unusually high star formation rate (SFR; e.g., submillimeter galaxies, SMGs) for which the contribution of SF winds in sustaining the outflow cannot be completely excluded \citep[e.g.,][]{Harrison2012}. 

Given that the critical coalescence/blow-out phase is expected to be very short ($<<500$ Myr; see, e.g., \citealt{Menci2008}), obscured and X--ray active (e.g., accretion on the BH is close to its maximum), the crucial point to study the feedback effects on the host galaxies is to select such objects as close as possible at the maximum of this phase. 
Large area X--ray surveys with the associated high-quality multiwavelength data provide the necessary tool for selecting these very rare sources. Luminous X--ray selected obscured AGN represent optimal targets for the search of objects where AGN feedback is expected to halt SF and to start ``cleaning'' out gas from the galaxy. In a previous work (\citealt{B10}, hereinafter B10) based on XMM-Newton observations of the COSMOS field (\citealt{Hasinger2007,Cappelluti2009}), we proposed a criterion for isolating the objects in the ``feedback'' phase on the basis of their observed red colors and high X--ray-to-optical and/or mid-infrared-to-optical flux ratios. A dedicated X-shooter campaign on a small subsample of such luminous sources at z$\sim$ 1.5 has convincingly shown that the proposed selection criteria are robust: the presence of outflowing material with velocities up to 1500 km s$^{-1}$ was indeed detected in six out of eight sources, confirming the efficiency of the selection in isolating such rare objects (\citealt{B14}, B14). 

Here we present a more in-depth study of the two brightest sources, XID2028 at z$=1.5927$ and XID5321 at z$=1.4702$, aimed at studying in more details the properties of the outflowing [OIII] ionized gas unveiled by B14, through slit-resolved spectroscopy, i.e. to map and locate the outflowing components outside of the central slit and determine its energetics. In addition, we show the presence of outflows also in the [OII]$\lambda\lambda$3726,3729 emission and, more interestingly, in the sodium (NaD$\lambda\lambda$5890,5896) and magnesium (MgII$\lambda\lambda$2796,2803 and MgI$\lambda$2853) absorption lines in both sources, corresponding to a neutral phase of the outflow.  
The paper is organized as follows. Section 2 presents the target properties; Section 3 presents the X-shooter data analysis and the results of the spectral fitting; Section 4 discusses additional proofs of outflowing material, investigating the presence of shifted components in the detected (MgII, MgI, and NaD) absorption and ([OII]) emission lines; Section 5 shows evidence of the AGN photoionization origin of the emission lines through rest-frame optical diagnostics; Section 6 discusses the energetic output associated to the outflow. Finally, Section 7 summarizes the main results and their implications. 
All the rest frame wavelengths are given in the air, as quoted in
http://www.sdss3.org/dr8/spectro/spectra.php.
Throughout the paper, we adopt the cosmological parameters $H_0=70$ km s$^{-1}$
Mpc$^{-1}$, $\Omega_m$=0.3 and $\Omega_{\Lambda}$=0.7 \citep{Spergel2003}.
%In quoting magnitudes, the AB system will be used, unless otherwise stated. 

\begin{figure}
\centering
\includegraphics[height=38mm]{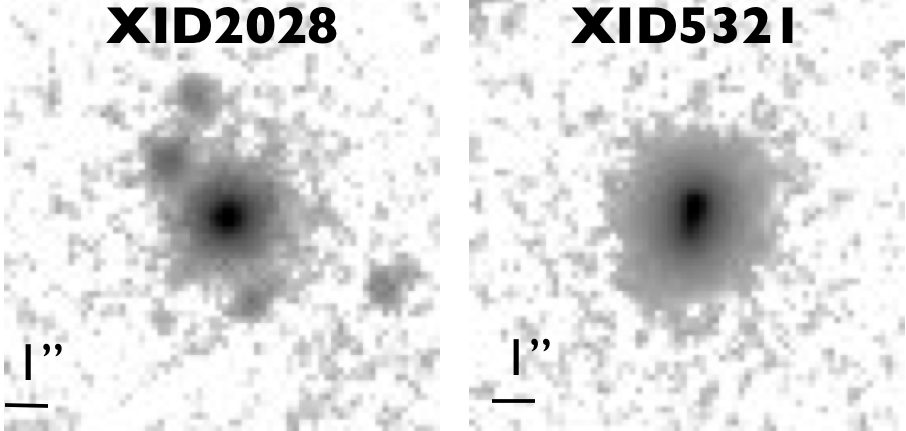}
\captionsetup{font={small}}
\caption{Cosmos CFHT/WIRCam H band images (10''x10'') of XID2028 (left) and XID5321 (right). Images are oriented with north up and east to the left.}
\label{newfig1}
\end{figure}

\section{Target properties}

We have completed an X-shooter near-infrared spectroscopy campaign on a sample of ten luminous X--ray selected obscured QSOs at z$\sim$ 1.5 from the XMM-COSMOS survey (B14). The targets were selected on the basis of their observed red colors and high X--ray-to-optical and/or mid-infrared-to-optical flux ratio (B10). 
Complete details about the target selection, the X-shooter observations, and data reduction are given in B14. Briefly, we recall here that all the sources for which the presence of outflowing gas has been detected share the following common properties: they are luminous AGN (with L$_{\rm bol}\sim10^{45-46.5}$ erg s$^{-1}$), X--ray absorbed ($N_H \sim10^{21-23}$ cm$^{-2}$) and they are  hosted in massive galaxies, with stellar masses $M_* = 10^{11-12}$ M$_\odot$.  
 
The two brightest X--ray targets, XID2028 and XID5321, with unabsorbed luminosity $log(L_{X})>45$ erg s$^{-1}$ in the 2-10 keV, are those detected at the highest S/N in the X-shooter NIR spectrum. 
These two sources are also extreme in the host galaxies properties: they have stellar masses on the order of 10$^{12}$ M$_\odot$ and exhibit substantial SFRs (SFR$\sim$250 M$_\odot$/yr, assuming a Chabrier IMF; see B14) as inferred from the Herschel PACS detection after a detailed spectral energy distribution (SED) fitting decomposition (B14, \citealt{Bongiorno2014}; see also below). 
%Most galaxies at z$\sim$ 1.5-2.5 are actively building up their stellar mass, forming stars at rates of 20-200 M$_\odot$ yr$^{-1}$ and sSFR= SFR/M$_* \sim$ 1-2 Gyr$^{-1}$ (i.e. the Main Sequence (MS) of star- forming galaxies; Rodighiero et al. 2011), with a fraction of galaxies (above MS) forming stars up to 10 times faster (starbursts like SMGs). 
The SFR observed in XID2028 and XID5321 are  consistent with those observed for the normal population of star forming galaxies at those redshifts and place the targets on the main sequence (\citealt{Whitaker2012,Rodighiero2011}; see also B14). %, i.e. they are not %actively building up their stellar mass (\citealt{Rodighiero2011}). 
Finally, they have BH masses on the order of $10^{9-10}\ M_\odot$, obtained as H$\alpha$ single-epoch virial estimates (\citealt{Bongiorno2014}). The H$\beta$ BLR emission lines are considerably extincted (see B14; Section \ref{secredd}), making these sources consistent with a type 1.8 nature. 

The radio flux measured at 1.4 GHz (observer frame) in the Very Large Array (VLA) observations of the COSMOS field (\citealt{Schinnerer2010}) are 102$\pm$20$\mu Jy$ and 180$\pm$24$\mu$Jy for XID2028 and XID5321, respectively. 
Assuming $S_{\nu}\propto \nu^{-0.7}$, %$S_{1.4GHz}=S_{obs}*(1/(1+z))^{-0.7}$, $S_{1.4GHz}(5321)=0.39 \pm 0.12 mJy$,  $S_{1.4GHz}(2028)=0.32 \pm 0.13 mJy$. \\
the radio luminosities are $L_{1.4GHz}(5321)=4.7\cdot 10^{24}$ W Hz$^{-1}$ and $L_{1.4GHz}(2028)=3.3\cdot 10^{24}$ W Hz$^{-1}$, corresponding to low ratios between the far-infrared and radio emission ($q_{IR} \sim 0.8$; \citealt{Ivison2010}); these values place the targets in the radio quiet class (see, e.g., \citealt{Bonzini2013}), guaranteeing marginal contribution from radio jets in the energetic of the systems.

Figure ~\ref{newfig1} shows the CFHT/WIRCam cutouts of the H band image of the two targets. %, where complexities and asymmetries are noticeable.
While for XID2028 we can firmly exclude the presence of two sources within 0.9'' (i.e., the size of our slit; see below) given the high resolution HST/ACS data available (see Figure 4 left, upper panel; see also B10, B14 and \citealt{Cresci2014}), the presence of two sources in XID5321, not resolved in our imaging data (for this source we lack ACS coverage), cannot be excluded. We will discuss this in more details in Section \ref{secsummary}.

%allow multiple slit resolved spectroscopy with large S/N to map the locate of the broad and shifted components out to the central slit. 

%X--ray obscured QSOs at $z \sim 1.5 $ from the XMM-COSMOS survey, expected to be caught in the transitioning phase from starburst to AGN dominated systems. 

\subsection{SED fitting}
\begin{figure*}
\begin{minipage}[t]{18cm}
\centering
\includegraphics[scale=0.5,angle=-90]{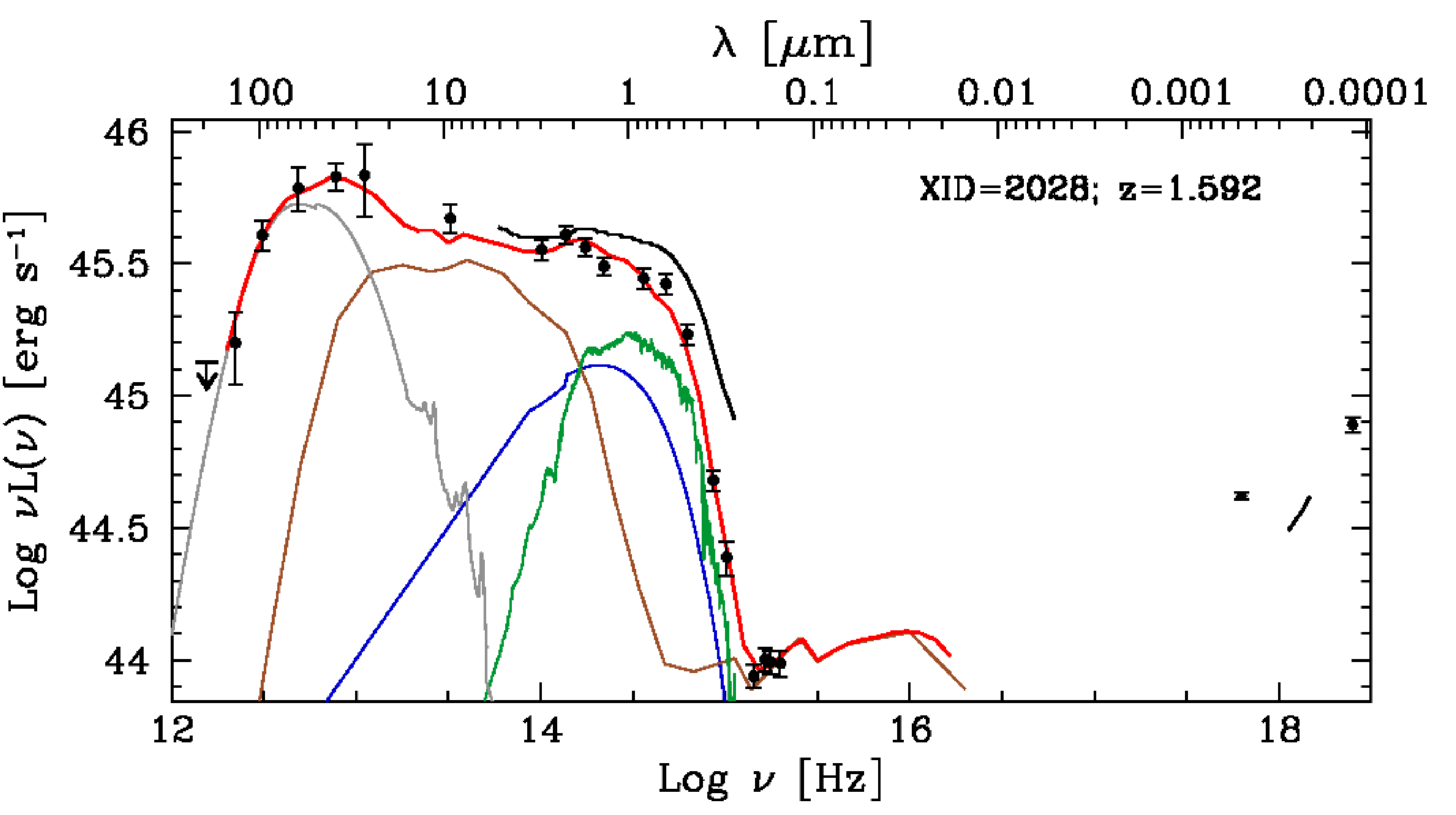}
\end{minipage}
\hspace{10mm}
\begin{minipage}{18cm}
\centering
\includegraphics[scale=0.5,angle=-90]{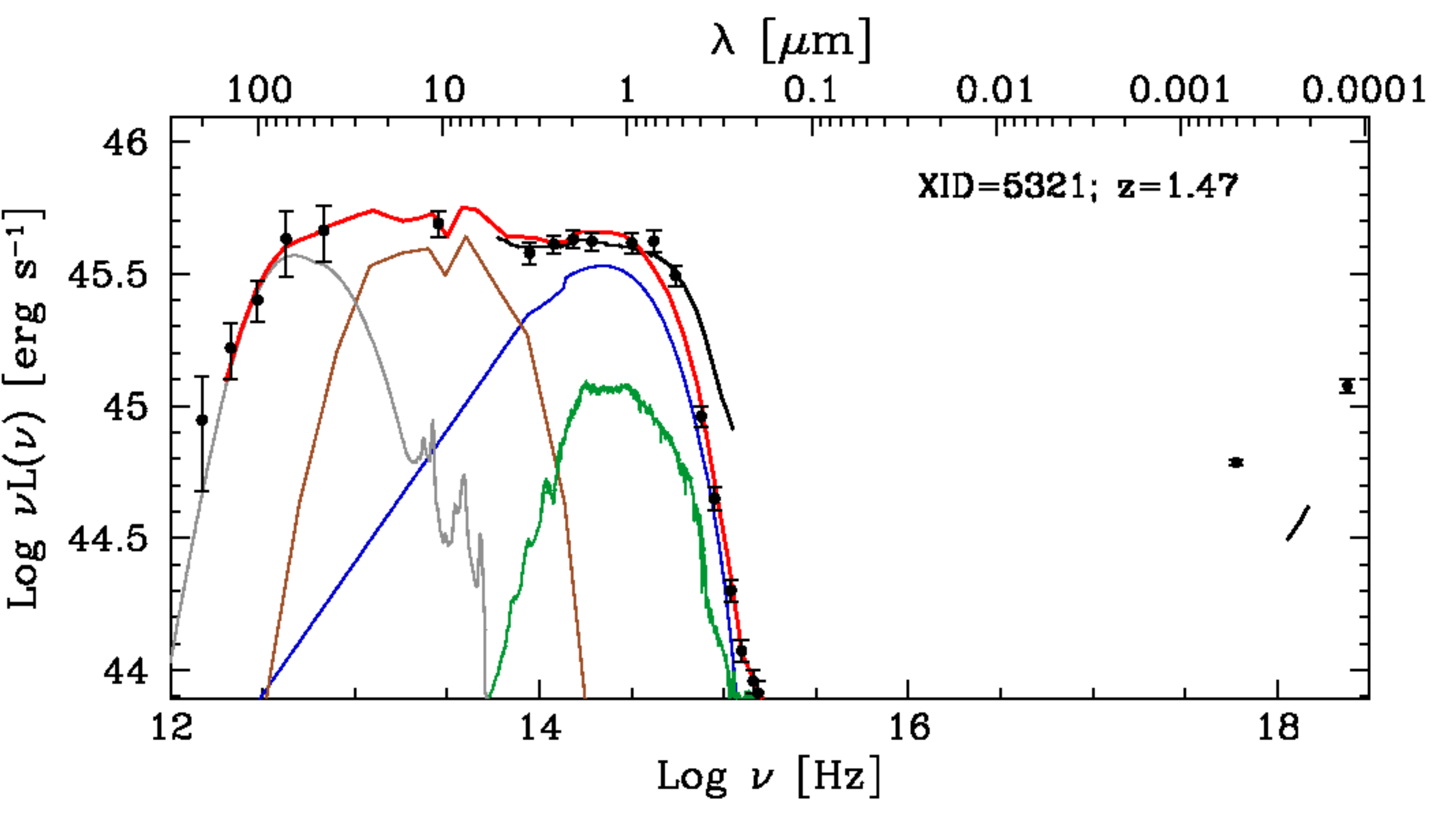}
\end{minipage}
\caption{SED fits of XID2028 and XID5321. Black circles are the observed photometry data. The grey, brown, blue and green lines correspond to the starburst, AGN torus and disk, and host-galaxy templates found as best fit solutions. The red lines represent the best fit SEDs. The black lines represent the mean SED computed by \citet{Lusso2011} for obscured AGN in the XMM-COSMOS survey at redshift z$=1.5$ and with similar bolometric luminosity.}
\label{sed}
\end{figure*}

The selected XMM-COSMOS sources benefit from a rich multiwavelength data-set 
with {\it complete} and {\it homogenized} ultraviolet to far-infrared (including Herschel data) to radio coverage, and extensive spectroscopic follow--up.  More details on the photometry collected for the counterparts of XMM-COSMOS can be found in B10, with the addition of the PACS datapoints (\citealt{Lutz2011}). The rest-frame SEDs of the two targets, previously presented in \citet{Lusso2012}, were reanalyzed using the SED-fitting code presented in \citet{Lusso2013} (see below for details), and are shown in Figure~\ref{sed}. In addition to the longer wavelength data, in each panel of Figure~\ref{sed} we also plot the X--ray datapoints in the soft and hard energy bands (black points at log $\nu \sim$ 18).

In obscured AGN the optical-UV nuclear emission is absorbed by the dusty torus and reprocessed in the IR. Therefore the average SED is characterized by an optical-UV luminosity much lower than that seen in unobscured AGN (e.g., the big blue bump (BBB), \citealt{Elvis1994}) and partly due to the host galaxy. The contribution of the AGN is maximum at MIR wavelengths (around 12 $\mu$m, \citealt{Gandhi2009}), while most of the emission at longer wavelengths can be ascribed to star formation only. 
The data points from far-infrared to the UV band  have therefore been fit with a combination of four components: 1) a stellar component to account for the host galaxy contribution (green curves); 2) a torus component to account for the reprocessed AGN emission in the MIR (brown curves); 3) a BBB component to account for the AGN accretion disk thermal emission (blue curves); 4) a starburst component to account for the on-going star formation (grey curves). The red lines represent the best fit SEDs. 
We notice that, given the quality of the data points (e.g., dense sampling, bright fluxes and small errors) and the distinctive shape of the SEDs, the decomposition in the different components is basically independent from the details of the fitting code, both in terms of templates used, and minimization procedure. In particular, we verified that all the physical parameters of interest (e.g., AGN luminosities, SFR, extinction) are solid and do vary only marginally (within 10-20\%) when other SED fitting codes developed in COSMOS are used instead \citep[e.g.,][]{Bongiorno2012,Delvecchio2014}. The only parameter which vary significantly (a factor of 2) is the stellar mass, contingent on the presence of the reddened BBB component in these targets.   
From these fits we infer reddening values E(B-V)=0.95 and 1.0 for the AGN components in XID5321 and XID2028, respectively. The extinctions are obtained by reddening the BBB template according to the Small Magellanic Clouds (SMC) law (\citealt{Prevot1984}). Overall, the AGN contribution to the bolometric output (in the range 1-1000 $\mu$m) is  $\sim60$\% for XID2028 and $\sim80$\% for XID5321.% This can be inferred in Figure~\ref{sed} from the fact that the SED in $\nu$F$_{\nu}$ scale is flat from the FIR to the MIR band. 
Accounting for the extinctions, we infer total bolometric luminosities of $L_{bol}\sim2\times10^{46}$ erg s$^{-1}$. Using the BH masses already mentioned, we obtain Eddington ratios of $L_{bol}/L_{Edd}\sim0.01-0.05$ (see B14).

Composite SEDs for a sample of Type 2 AGN in XMM-COSMOS are presented in \citet{Lusso2011} for different bins of redshifts and luminosities. In each panel of Figure~\ref{sed} we have plotted the composite SED constructed in a similar range of luminosity of our targets (log $L_{\rm 8\mu m}\sim44.3-45.7$, which roughly corresponds to log $L_{\rm bol}\sim45.3-46.7$). %The average Type 2 SED is normalized to our data points at 1 $\mu$m.
From the comparison of the SED of our targets and the average SED of X--ray selected sources at the same luminosity, we note that the X--ray datapoints of  XID2028 and XID5321 are above the average SED by $\sim0.5$ dex.
This confirms that these 2 targets are X--ray loud (as per their selection on the basis of the high X/O ratio) but optically weak (without a clear view of the accretion disc emission). %This fits with a scenario where these objects are getting to the maximum of their accretion phase. This is consistent with the accretion rate observed for the two targets (L/L$_{Edd}\sim0.01-0.05$; see B14).
Indeed, the flat shape between 1-10 $\mu$m %suggests that these sources are probably dominated by the AGN emission at those wavelengths 
and the presence of BLR lines in the NIR X-shooter spectra, are all consistent with a transition from type 2 to type 1 through a reddened type 1 phase.

%-----------------------------------------

\begin{table}
\begin{minipage}[t]{10cm}
\footnotesize
\centering
\caption{Observations log}
\begin{tabular}{lcccc}
XID & start date &  Seeing \footnote{ Delivered seeing corrected by airmass} & expo & P.A.\\
\toprule
2028 & 2013-02-09 & 0.92 & 3600 & 90\\
& 2013-02-09 &  & 3378 & \\
&2008-01-08\footnote{Keck/DEIMOS}  & 1 & 9600 & \\ \hline
5321 & 2013-02-10 &  0.74 & 3600 & 135 \\
& 2013-02-10 &  & 3378 & \\ \toprule
\end{tabular} 
%\footnote{ Delivered seeing corrected by airmass}
\end{minipage}
\end{table}

%-----------------------------------------

\subsection{X--ray spectra}\label{secX}

For both sources medium-deep X--ray spectra are available from the XMM-COSMOS survey (\citealt{Mainieri2011}). X--ray spectra were extracted using standard techniques and calibration files, following the procedures described in details in \citet{Mainieri2011} and \citet{Lanzuisi2014}.%: briefly, 
%\footnote{Yaxx is a Perl code, developed by T. Aldcroft at the Harvard Smithsonian Center for Astrophysics, which automatically deals with the extraction and fitting of sources detected in large Chandra and XMM-Newton surveys, such as ChaMPs (Kim et al. 2004) and COSMOS. Yaxx makes use of  XMM-Newton Scientific Analysis Software (SAS6) tools for the spectra extraction and ARF and RMF creation with CALDB 4.1.2 calibration files.} 
%we extracted source and background counts, and  created response matrices, ARF and RMF, for each source and in in each observation separately. Then 
The counts from the same camera (MOS, pn) are merged together, and the responses are created for each camera. % with the tools \emph{addarf} and \emph{addrmf} (see \citealt{Lanzuisi2013} for more details). 
The total number of net counts used in the spectral fitting sums up to $\sim1500$ for both sources.

\begin{figure*}
\centering
\includegraphics[scale=0.4]{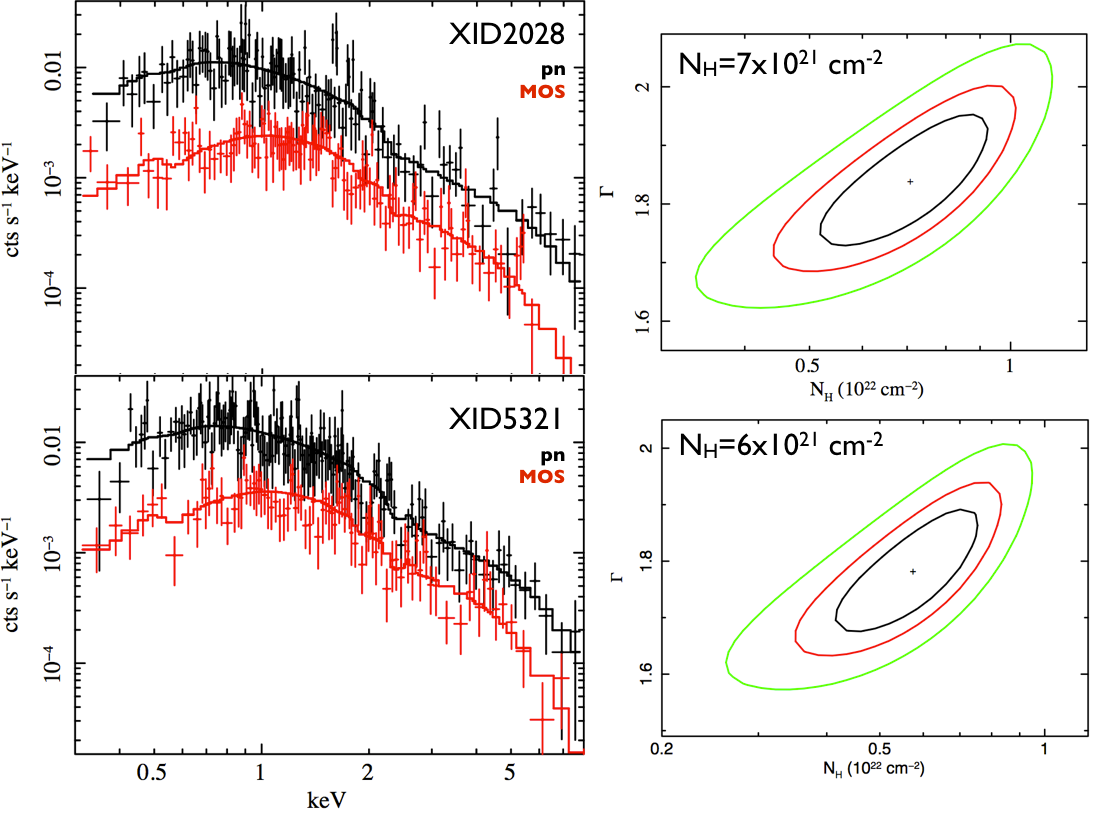}
\caption{X--ray spectra of XID2028 (top, left) and XID5321 (bottom, left). The different datasets used (XMM pn and MOS), are labeled in different colors. The best fit model is also shown as a solid line for each dataset. On the right are shown the combined constraints on N$_{H}$ and the spectral slope, with confidence contours at 68, 95 and 99\% level. }
\label{Xray}
\end{figure*}

The joint fit to the pn and MOS XMM spectra are presented in Figure~\ref{Xray} (left panel). An absorbed power law provides a good fit over the $\sim$ 0.3--8 keV energy range. For both sources the best fit intrinsic slope and absorption column density are similar: $\Gamma\simeq1.84\pm0.12$,  $N_H\sim7.1\pm0.2\times$ 10$^{21}$ cm$^{-2}$ for XID2028, and $\Gamma\simeq1.78\pm0.12$,  $N_H\sim5.8\pm0.2\times$ 10$^{21}$ cm$^{-2}$ for XID5321. 
The $N_H$ versus $\Gamma$ confidence contours (68, 90 and 99\%) are shown in Figure ~\ref{Xray} (right panels). There is no evidence of iron K$\alpha$ emission with 90\% upper limits of 100 eV and 250 eV (rest--frame) in XID2028 and XID5321, respectively. The addition of a Compton reflection component subtending a 2$\pi$ angle at the continuum source marginally improves the fits ($\Delta\chi^2 \sim$2). The intrinsic spectral slope gets steeper $\Gamma\sim$ 1.93--1.99 and the intrinsic absorption higher $N_H \sim$ 7--8 $\times$ 10$^{21}$ cm$^{-2}$ respectively. 
The moderate, but significant intrinsic X--ray absorption argues against a pure Type 1 nature, while it is fully consistent with the Type 1.8 classification obtained by the near infrared X-shooter spectrum.
% MOVED TO DISCUSSION: 
%The observed intrinsic absorption could be consistent with the disk of the host galaxy, if seen edge on, or with some circum-nuclear material such as a low column density of the torus or the outer parts of a thicker one. The optical reddening ($E(B-V) \simeq$ 0.95 and 1.0 for XID5321 and XID2028 respectively) is fully consistent with the cold gas absorption measured in X--rays for a Galactic dust to gas ratio. 
The moderate obscuration observed in the X--rays coupled with the Eddington ratio ($\lambda_{\rm Edd}\sim1-5$\%) places these objects in the region of ``shortlived clouds'' presented in \citet{Raimundo2010}, where outflows or transient absorption are expected to happen.
% In fact, 
The optical reddening is fully consistent with the cold gas absorption measured in X--rays for a Galactic dust to gas ratio (\citealt{Maiolino2001}). %ok ma non valeva per E(B-V) da BLR?
\footnote{ $N_H$ determines to which wavelengths the medium is opaque and to which wavelengths the medium is transparent, where a medium with $N_H>10^{21} cm^{-2}$ will obscure the UV emission, a medium with $N_H>10^{22}$ will obscure the broad H$\alpha$ line and the 3 $\mu$m emission, and a medium with $N_H>10^{23}$ will obscure the 10 $\mu$m emission. All the numbers assume a Galactic dust composition and dust to gas ratio. }
 The observed intrinsic absorption could therefore be consistent with material in the host galaxy, or with some circum-nuclear material such as a low column density torus or the outer parts of a thicker one.
 %The observed E(B-V) is consistent with the Is this a proof that the obscuration is on larger scales?
 % Given that these more luminous sources are also usually associated to host galaxies with very high SFR (e.g., SFR can be as high as 4000 M$_{\odot}$ yr$^{-1}$ in J1234, Banerji et al. 2014),
 %
 % J1237: Bongiorno+14: NH=3e22, a factor of 10 $\sim$lower Lx; similar SFR
 %
 
\section{NIR slit-resolved spectroscopy analysis}\label{secoutflowprop}

XID2028 and XID5321 have been observed with the X-shooter spectrograph (\citealt{Dodorico2006,Vernet2011}) on the ESO VLT-UT2 (Kueyen) during the nights of February 8-10, 2013 (ID: 092A.0830, PI: Brusa).
As presented in B14,  we revealed the presence of outflowing components associated to the most prominent Narrow Line Region (NLR) emission lines in the H$\beta$-[OIII] and H$\alpha$-NII regions, with velocity shifts $V_S$ with respect to the systemic features, on the order of  $\left | V_S \right | \sim$ 350 - 450 km s$^{-1}$, in the nuclear spectra of both sources on an aperture of $\sim 1''$. XID2028 shows a blueshifted component, while XID5321 shows a less common redshifted component (see also \citealt{Villar2011}, \citealt{Harrison2012}, \citealt{Bae2014} for other examples of [OIII] lines with redshifted broad wings). 

Here we analyze in more detail the X-shooter spectra of the two sources, with the aim of determining more accurate estimates of the key parameters (spatial extension, outflow velocity and flux) related to the outflowing winds. 

\subsection{Spatial extension}\label{secextension}
\begin{figure*}[!]
\centering
\includegraphics[scale=0.47]{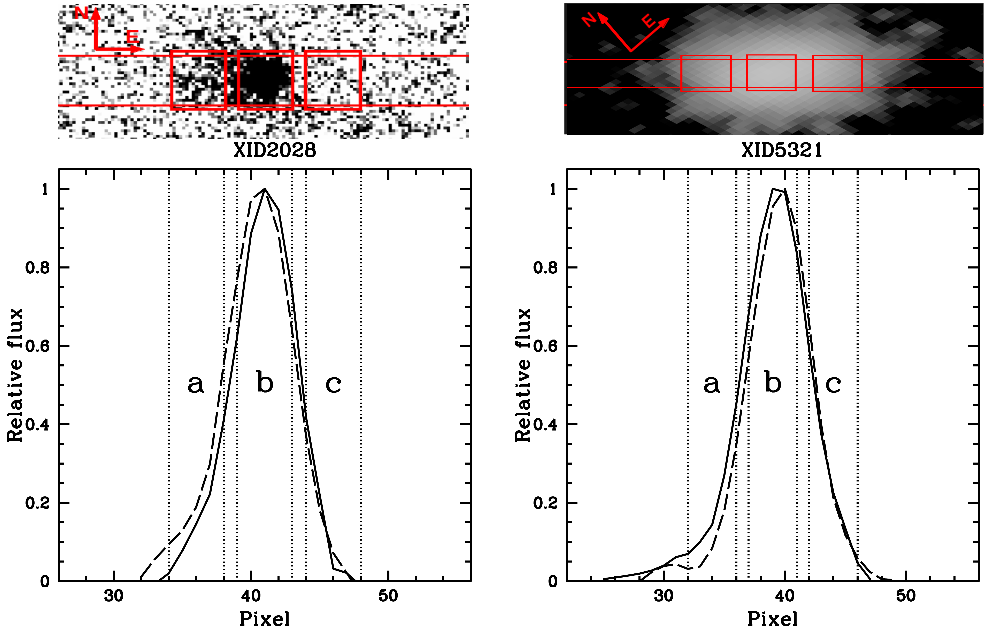}
%\captionsetup{font={small}}
\caption{({\it lower panels}) Spatial profiles of the 2D spectrum of XID2028 (left) and XID5321 (right) of the continuum flux (solid curves) and in the proximity of [OIII]$\lambda$5007 emission line (long dashed curves). Both continuum and [OIII]$\lambda$5007 wavelength range profiles are normalized to unity. The dotted lines
shows the demarcations of the three regions from which the apertures were extracted (denoted with $a$, $b$, and $c$). ({\it upper panels}) HST/ACS F814 image of XID2028 (left) and J-band UltraVista image of XID5321 (right) showing the position and orientation of the 0.9'' slit and the spatial locations of the three apertures.}
\label{profilisp}
\end{figure*}

% (MOVE IN THE CONCLUSION?) In order to trace the spatial extent of the outflows in our 2 brigthest X-shooter targets we performed spatially resolved spectroscopy. 
The analysis presented here is based on spectra extracted from three different spatial regions ($a$, $b$, $c$) of each system along the slit (see Figure \ref{profilisp}). 
For each source, the $(b)$ region is centered on the nucleus as determined from the peak of the spatial profile of the continuum emission (see below), and covers a  projected region of 0.85x0.90'', i.e. $\sim$ 7.3 x 7.7 kpc. The $(a)$ and $(c)$ apertures, instead, cover the same projected regions in size, but are extracted from two regions offset from the nucleus. 
The slit position of the sources and the spatial location of each aperture are shown in the upper panels of Figure ~\ref{profilisp}.%, overlaid on the HST/ACS F814 (Advanced Camera for Survey) cutout of XID2028 (left) and on the J-band UltraVista image of XID5321 (right). %The last one is the image with the best resolution available in COSMOS for this source ({\bf try also zpp image, Subaru}).

The spatial profiles along the slit of the 2D spectra are shown in Figure ~\ref{profilisp} (bottom panels). 
The dotted lines demark the regions indicating the location of the three apertures described above, as labeled. The {\it solid curves} show the spatial profile of the continuum flux, obtained by averaging  profiles extracted at many line-free regions along the entire dispersion axis; the {\it dashed curves} show the spatial profile of the emission in the [OIII]$\lambda$5007 line. For XID2028, the different profiles of the solid and dashed curves represent the first evidence that complex narrow line kinematics is in place. %for these sources. 
The small displacement between the spatial profiles of XID5321 is instead due to a slight tilt along the dispersion axis. Hence, a possible error in the determination of the peak of the spatial profile of the continuum flux, and therefore of the location of the nuclear region of XID5321, could be present. However, it does not significantly change the results in the following analysis, given that 1 pixel gaps between the three regions are present.

\begin{figure*}[!]
\centering
\includegraphics[scale=0.37]{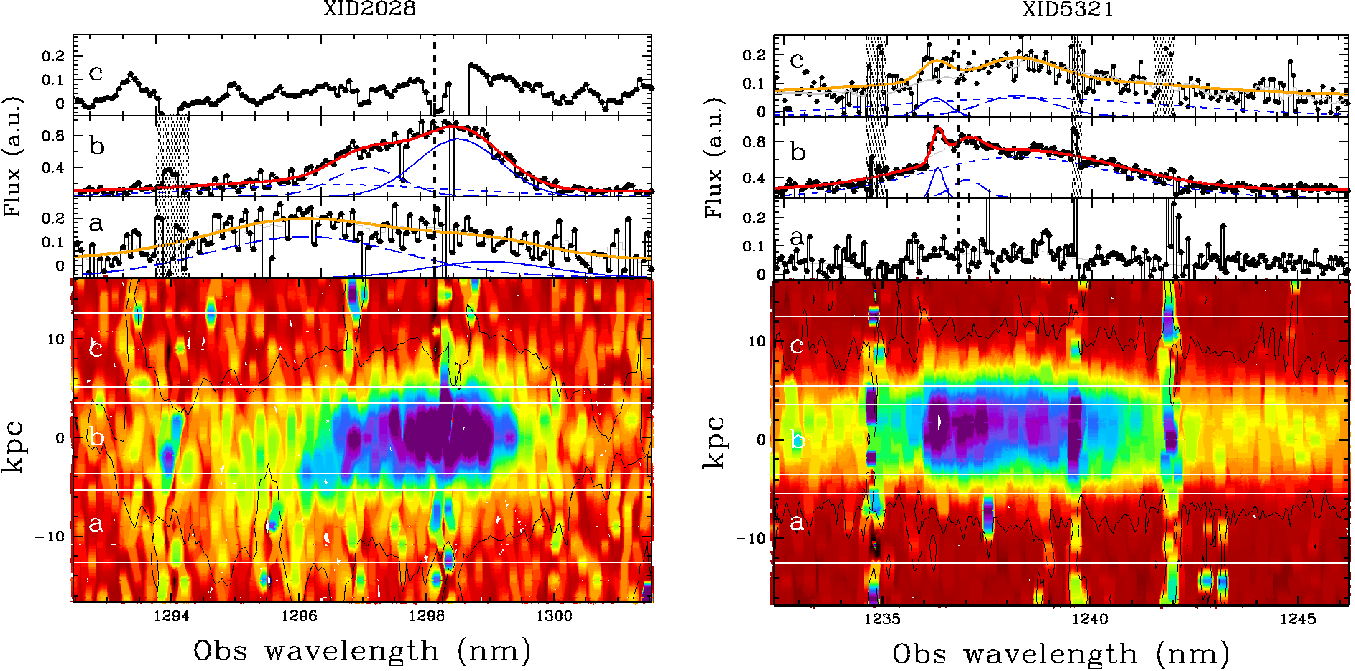}
%\captionsetup{font={small}}
\caption{XID2028 (left) and XID5321 (right) X-Shooter spectra of the regions $a$, $b$, $c$, centered on [OIII]$\lambda$5007. % region for XID2028 (left) and XID5321 (right), in three different apertures along the slit: the central one on a $\sim$ 4 kpc radius from the AGN, and the two others from $\sim$ 6 to 12 kpc in opposite directions along the slit. 
The red and orange lines indicate the best fit solutions that reproduce the line profiles according to the non-parametric approach (see Section \ref{secnonparametric}). The Gaussian components are shown with arbitrary normalization in order to ease the visualization. The dotted lines mark the wavelength of [OIII]$\lambda$5007 at the systemic redshift determined by B14. The lower panels show the 2D spectra, indicating the apertures used to extract the 1D spectra seen in the upper panels. Red to blue colors represent increasing flux. Extended vertical structures, like the one at $\sim$ 1242 nm (right), delineate strong sky features and are indicated in the 1D spectra as shaded areas.} %The high velocity blueshifted (XID2028, aperture $a$) and redshifted (XID5321, aperture $c$) tail tracing the outflow extend up to 10-12 kpc in one of the two off-center positions.}
%Blue and magenta lines mark the different components used to reproduce the line profiles (in red) according to the second fit approach.}
\label{aperture2D}
\end{figure*}
%profili5321new.sm
%

Figure ~\ref{aperture2D} shows the  spectra extracted from each region, indicated with letters ($a$, $b$, $c$),  centered on [OIII]$\lambda$5007. In the figure, solid lines represent the fits that best reproduce the line profiles; details are given in the next Sections. 
For both sources, significant signal is detected only in one off-nuclear aperture (aperture ($a$) for XID2028, ($c$) for XID5321).  We note that seeing conditions were $\sim$ 0.8'', hence the extension seen ($\gtrsim$ 2 $\times$ seeing) is fully resolved and is not due to seeing effects.  
Figure  ~\ref{aperture2D}  shows that complex multicomponent profiles in the [OIII]$\lambda$5007 line are detected not only in the nuclear region (covering up to 4-5 kpc, see B14), but out to a distance of R $\approx$ 10-12 kpc, therefore extending considerably over the host galaxy. %A comparison with the results of \citet{Cresci2014} may be carried for XID2028: 
SINFONI observations of XID2028 (\citealt{Cresci2014}) confirm that extended [OIII] emission, roughly orientated along our X-Shooter slit, is extending up to 13 kpc from the nucleus. We therefore assume that the outflow in each source extends out to $\approx 11$ kpc from the central black hole. % for both sources.  

% (MOVE TO DISCUSSION?) The observed extent, although obtained through NIR integrated spectra, is comparable to those measured from IFU data in high-z galaxies (.e.g Harrison et al. 2012, $\sim5-20$ kpc) and significantly larger than what has been observed ($\sim3$ kpc) in the very luminous QSO presented in Cano-Diaz et al. (2012). 

\subsection{Velocities and fluxes}
To determine the dynamics and the outflow properties, we proceed using two approaches: a simultaneous line fitting and a non-parametric analysis.
The two approaches are justified by the fact that with the simultaneous line fitting we can have a better estimate of the unperturbed and outflowing gas fluxes separately, while with the non-parametric analysis we can infer more accurate outflow velocity measurements.
In both cases, the spectra were shifted to the rest frame using the redshift found by B14 and the continuum flux in the vicinity of emission lines was fit with a power-law model.

\subsubsection{Simultaneous modeling the H$\beta$ $+$ [OIII] and H$\alpha$ $+$ [NII] line complexes}\label{secsimultanei}

The first approach is the one described in B14, applied to three different apertures along the slit. In this case, to derive the outflow properties we used the constraints in accordance with atomic physics (i.e., doublet flux ratios, wavelength separations; for more details see B14). Briefly, we fit simultaneously the two regions (H$\beta +$ [OIII] and H$\alpha+$ [NII]$\lambda\lambda$6548,6583+[SII]$\lambda\lambda$6716,6731) %, over two power-law fits to the local continua, 
with three sets of Gaussian profiles in the nuclear apertures: two sets of narrow Gaussian lines, with FWHM $\lesssim 500$ km s$^{-1}$ and FWHM $\gtrsim 500$ km s$^{-1}$ respectively, for each detectable emission line, %(namely:  H$\alpha$ and H$\beta$, the [OIII], [NII] and [SII] doublets),
and a broad (FWHM $>2000$ km s$^{-1}$) Gaussian or Lorentzian line to account for the presence of the H$\alpha$ and H$\beta$ emission originated in the BLR.
In the off-nuclear apertures, where we do not expect to see BLR emission lines, and the S/N is lower, %especially to try to deblend the components in the [NII]+H$\alpha$ profile (requiring six Gaussian lines), 
only one set of Gaussian profiles is used, with no limit to the FWHM.

\begin{figure*}
\centering
\includegraphics[scale=0.53,angle=-90]{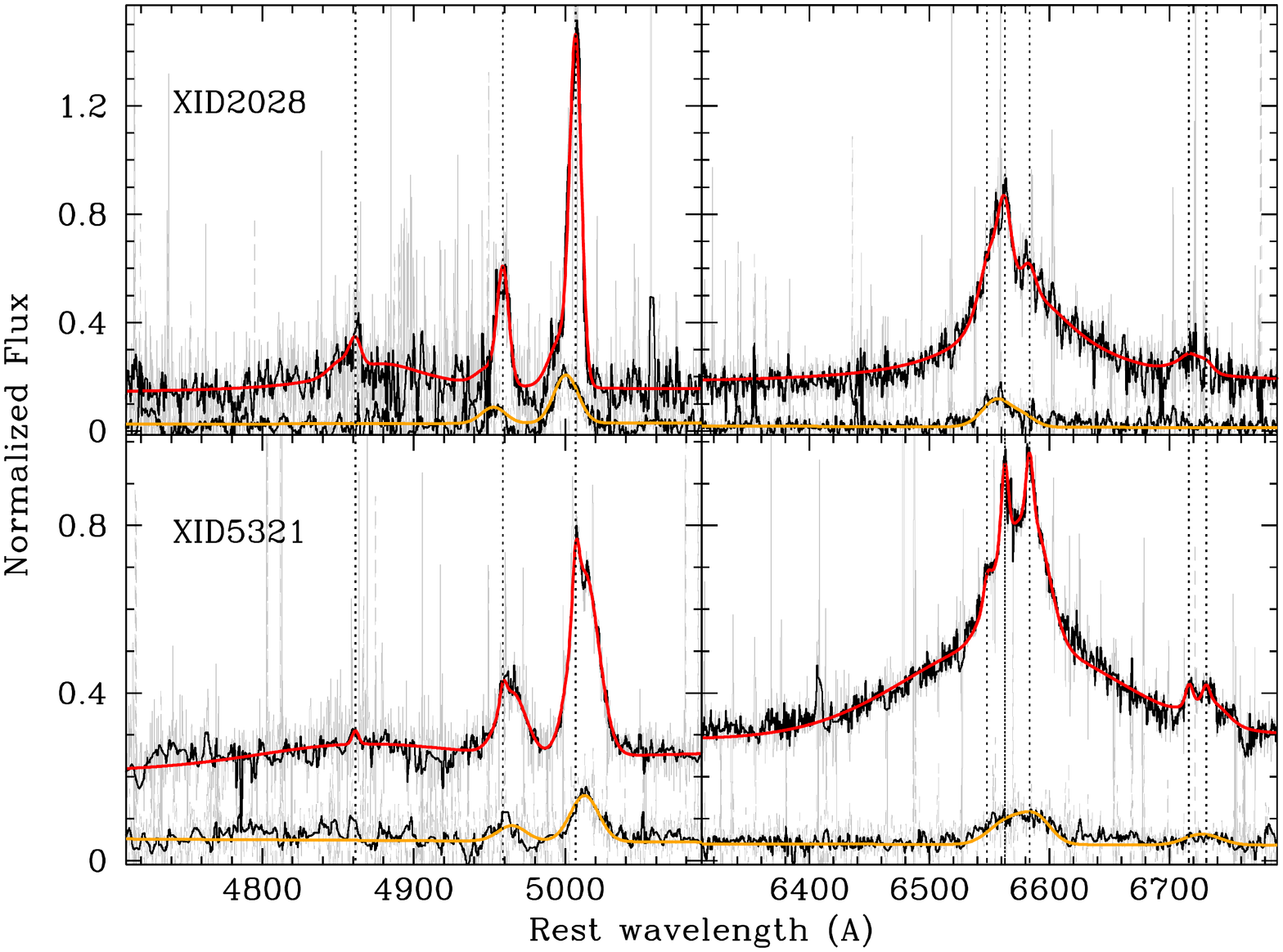}
%\captionsetup{font={small}}
\caption{Spectra of the two quasars normalized so that the intensity of the strongest emission line of the nuclear aperture is unity. For clarity, in the figure the nuclear spectrum of XID2028 is shifted vertically in order to facilitate a visual inspection of the apertures. The grey and black lines indicate the original and the cleaned (of cosmic ray hits, bad pixels and telluric lines) spectra. The red and orange lines indicate the best fit solutions that reproduce the line profiles of the central and the lateral apertures according to the simultaneous fit approach. The dotted lines mark the wavelengths of H$\beta$, [OIII]$\lambda$4959, [OIII]$\lambda$5007 (left) and [NII]$\lambda$6548, H$\alpha$,[NII]$\lambda$6581, and the [SII] doublet (right), from left to right, at the systemic velocity. }
\label{fitsimultanei}
\end{figure*}

Figure ~\ref{fitsimultanei} shows the best fit solution of our spectra, as in Fig. 6 of B14, for the apertures $a$ and $b$ for XID2028 and $b$ and $c$ for XID5321. It turns out, especially for the [OIII] emission lines, that the profiles of the nuclear regions (red curves in the figure) are broad and asymmetric. We do not see the same effect for the other lines, H$\alpha$, H$\beta$ and the forbidden [NII] and [SII], because they are blended with the BLR component and/or of lower S/N. 
%The same considerations come from the inspection of the Figure ~\ref{profilivel}, relatively to the results of the first method.
%We obtained velocity shift $V_S \sim$  350 km s$^{-1}$.

%Although in this second approach different emission lines that could come from different regions and with different kinematics are fitted simultaneously, this method undoubtedly brings many benefits. 
Using all the constraints due to the atomic physics, and therefore minimizing the degeneracy (especially in the [NII]+H$\alpha$ profile), this approach 
allows us to show, in the nuclear regions, 
%{\bf This approach has shown 
the presence of one component associated with overall narrow profiles (FWHM $\sim$ 300-500 km s$^{-1}$), likely associated with the systemic %unperturbed
NLR, and of another component with FWHM $\sim$ 1300-1600 km s$^{-1}$ and shifted with respect to the narrower ($V_S \sim$ 340 km s$^{-1}$), associated to outflowing emitting gas. Moreover, it allows us to study the physical conditions of the ionizing gas of these two components using the line flux ratios (see Section ~\ref{seclineratio}). 
The profiles in the off-nuclear regions (orange curves in Figure ~\ref{fitsimultanei}) appear broader and shifted with respect to those determined in the nuclear regions (see Section \ref{secresfit}), and are likely associated to outflowing emitting gas. %This is a confirmation of what is seen in Figure~\ref{profilivel}, extended also to the other emission lines.

In the following, we will refer to the ``outflow'' components as the Gaussian profiles with FWHM $\gg 500$ km s$^{-1}$ of the best fit solution of this approach, not to be confused with the systemic narrow (FWHM $\lesssim 500$ km s$^{-1}$) profiles, likely associated to the NLR.

\subsubsection{Non-parametric analysis}\label{secnonparametric}

We used the non-parametric analysis for the only [OIII]$\lambda$5007 emission line.
Non-parametric measures do not depend strongly on the specific fitting procedure and on the physical interpretation of any of the parameters of the individual Gaussian profile, pointing to obtain a noiseless approximation to the emission line profile. 
%The first approach is based on the following:
Hence, once a local continuum was subtracted, we fit the [OIII] with as many Gaussian profiles as needed (3 at most) in order to reproduce the overall shape of the line profile.% In the fitting procedure, we used the constraints in ac with atomic physics (i.e. doublet flux ratios, wavelength separations; for more details see B14). 

The best fit solutions are shown in Figure ~\ref{aperture2D} for the different apertures along the slits. Gaussian components with widths up to $\sim$ 1000 km s$^{-1}$ are required, especially in the off-nuclear apertures.
In particular, the [OIII] line of the nuclear aperture of XID5321 presents a peculiar profile, with different peaks. Two out of the three Gaussian lines that reproduce the total profile have width smaller than 500 km s$^{-1}$, and could be attributed to the NLR and star forming regions. Alternatively, they can be originating in two distinct systems (e.g., merging galaxies) not resolved by our imaging data.

Once the profile has been accurately reproduced, we used non-parametric measurements of the [OIII] emission line profiles \citep[e.g.,][]{Zakamska2014,Liu2013,Rupke2013} carried out by measuring velocity $v$ at which a given fraction of the line flux is collected, using the cumulative flux function $F(v)=\int_{-\infty}^{v} F_v(v')\, dv'$. The position of $v=0$ of the cumulative flux is determined using the systemic redshift found by B14.
%{\bf Fitting the only [OIII] emission lines we were able to follow the line profiles even in the lateral apertures with 2-3 Gaussian lines (see Figure \ref{aperture2D})}. 
Using the best fit solutions and following the prescription indicated by \citet{Zakamska2014}, we estimated for the total [OIII] profiles the following parameters:
%(of no more than 3 Gaussians)
\begin{enumerate}[label=(\roman*)]
%\item 
%The line-of-sight velocity, represented by the median velocity, $v50$, i.e. the velocity that bisects the total area underneath of the line profile, so that $F(v50)=0.5F(v_{\infty})$, likely dominated by galaxy kinematics (Harrison et al. 2014);\\
\item 
The line width $w80$, the width comprising 80\% of the flux, that for a Gaussian profile is very close to the FWHM value. It is defined as the difference between the velocity at 90\% ($v90$) and 10\% ($v10$) of the cumulative flux, respectively;\\
%\item 
%A ``relative asymmetry'' parameters A, defined as $((v90-v50)-(v50-v10))/w80$, related to the standard skewness. $A$ is positive/negative if the profile has a strong redshifted/blueshifted wing, and is null for a symmetric profile;\\
\item 
The velocity offset of the broad underlying wings, $\Delta V$, measured as $(v05+v95)/2$, with $v95$ and $v05$ defined as above;\\
\item 
 A maximum velocity parameter $v_{max}$, defined as $v02$ when blue prominent broad wings are present, or as $v98$ when red ones are, on the contrary, present.
\end{enumerate}
In contrast to $v_{max}$, values of $w80$ and $\Delta V$ include only differences between velocities and do not depend on the accurate determination of the systemic velocity. A possible residual error in the determination of the systemic velocity, however, may produce a variation of at most few tens of km s$^{-1}$ in the $v_{max}$ value, corresponding to variation of a few \% (see below). 
Figure ~\ref{profilivel} shows the comparison between [OIII] velocity profiles for the nuclear (upper panels, red curves) and the off-nuclear regions (lower panels), with the non-parametric measurements of the best fit labeled. The profiles in the lateral apertures appear broad and shifted with respect to the systemic velocity determined in the nuclear regions, thus confirming the spatial extent of the outflows. In particular, $v_{max}$ increases by $\sim$ 200 %and 650 
km s$^{-1}$ going from the nuclear to the off-nuclear apertures, for 
%XID2028 and XID5321 respectively.
both sources.
%Dissimilarities between the velocity offsets $V_S$ and the non-parametric $\Delta V$ could be attibute to the degeneracy in the simultaneous fitting procedure with two components.
The [OIII]$\lambda$5007 of XID5321 shows, in both apertures, also a blueshifted wing, although it is much less extended than the red wing.

We stress here that the use of a larger sets of Gaussian profiles permits a better approximation to the emission line profile and more accurate velocity measurements, but increases the degeneracy between the Gaussian components, and thus does not allow a flux estimation for the NLR and outflowing emitting gas separately, in contrast to the simultaneous fit approach. Thus both approaches are crucial for our analysis.

\subsection{Results}\label{secresfit}

We  fit the Gaussian profiles using a fortran code implementing the Minuit package (\citealt{James1975}). In order to estimate errors associated to our measurements, we used Monte Carlo simulations. For each fit parameters, we collected 1000 mock spectra using the best-fit final models (red and orange curves in Figure ~\ref{aperture2D} and ~\ref{fitsimultanei}), added Gaussian random noise (based on the standard deviation of the corresponding local continuum), and fit them. The errors were calculated by taking the range that contains 68.3\% of values evaluated from the obtained distributions for each component/ non-parametric measurement.
%1000 Monte Carlo random spectra that were generated using the final models, for both the approaches: we collected 1000 mock emission line profiles adding random  noise (based on the standard deviation of the corresponding local continuum) to the final models, and we fitted them. From these we obtained the distributions for each component/non-parametric measurement. %La necessita' della terza componente e' definita dalla minimizzazione quando il fit e' facile (OII 5395)
We also note that since line profiles are non-Gaussian and much broader than the spectral resolution (see Section \ref{secabsorp}), we did not correct the observed profiles for instrumental effects and report all values as measured. 

In Table ~\ref{tvel} we  quote the v$_{max}$, the line widths $w80$ and the velocity shifts $\Delta V$ as obtained from the non-parametric analysis, and the FWHM of the outflow components and the velocity shift $V_S$ from the simultaneous fitting, all values which are used as estimator of velocity of the outflow in the literature \citep[e.g.,][]{CanoDiaz2012,Harrison2012,Harrison2014,Westmoquette2012}.
%The results obtained from the emission lines fit with the two methods are presented in Table ~\ref{tvel}; 
We underline that simultaneous fit results, i.e. the first three columns, regard all the emission lines and not only the [OIII]$\lambda$5007 emission.

The velocity estimators obtained in the off-nuclear regions have all very high values (e.g., $v_{max}\approx 1530$ km s$^{-1}$ and $\approx 1950$ km s$^{-1}$ for XID2028 and XID5321 respectively).
Overall, all the velocity estimators are greater in the lateral apertures by up to 45\%.  
Differences between the velocity offsets $V_S$ and the non-parametric $\Delta V$ estimators are due to the fact that $\Delta V$ depends strongly on the relative flux contributions of different components of the line profile. %could be attributed to the necessity, in the simultaneous fitting procedure, of using only one Gaussian component: an inspection of Figure ~\ref{aperture2D} suggests that one component profile, in order to minimize the residuals, tends to center the Gaussian nearest to the systemic and therefore to underestimate the shift.

%The blue and magenta lines in Fig2? mark the different Gaussian components used to reproduce the line profiles (the red lines) on which the non parametric measurements are carried. 

\begin{figure*}
\begin{minipage}[b]{9.0cm}
\centering
\includegraphics[width=7cm,height=7cm]{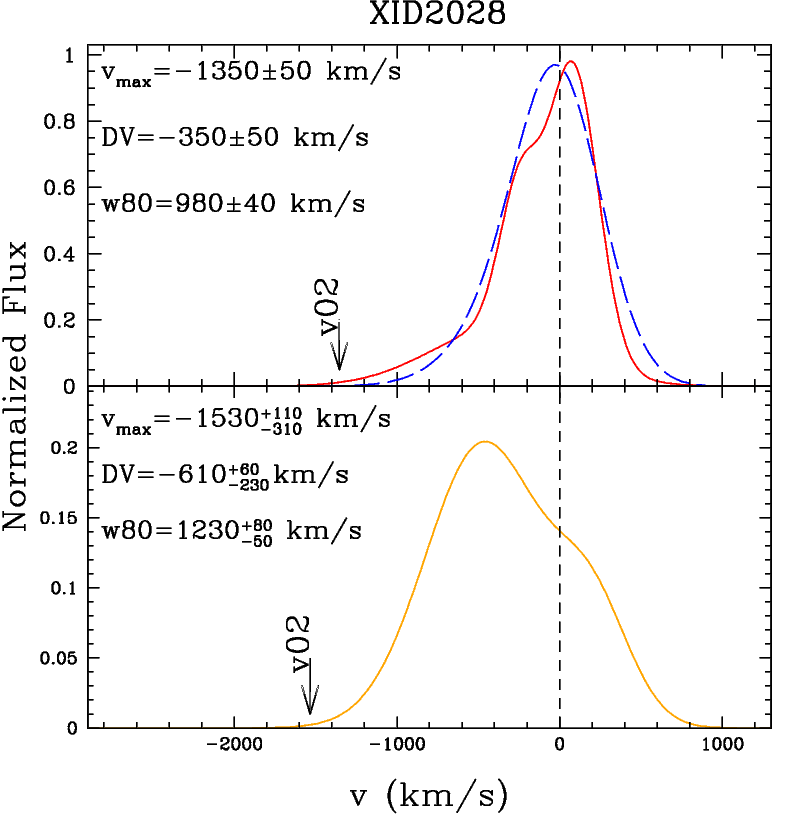}
\end{minipage}
\hspace{10mm}
\begin{minipage}[b]{9.0cm}
\centering
\includegraphics[width=7cm,height=7cm]{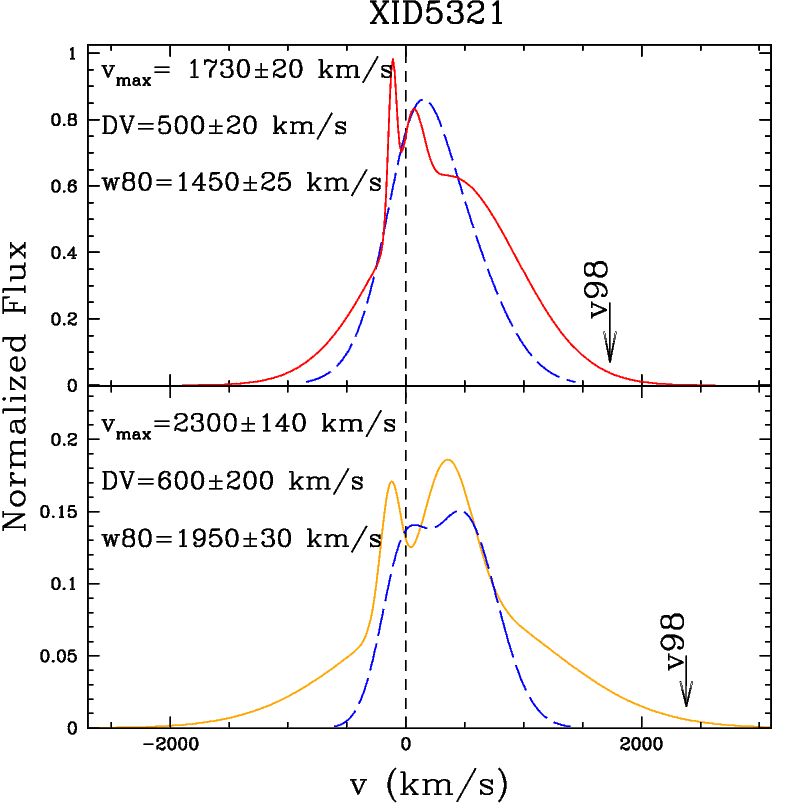}
\end{minipage}
\caption{Comparison between [OIII]$\lambda$5007 velocity profiles: red curves for the nuclear apertures in the upper panels, orange curves for the off-nuclear apertures retaining the outflow emission in the lower panel. The non parametric measurements of the [OIII] fit are labeled. We also overplot in long-dashed blue the results for [OII] emission lines fit (see Section \ref{secOII}). The [OII] profiles are renormalized in order to facilitate a comparison with the [OIII]. The dashed lines mark the positions chosen as zero velocity (the systemics for [OIII], and the mean wavelength of the [OII] doublet), the arrows mark the 98th and 2th percentiles indicating the $v_{max}$ velocities.}
\label{profilivel}
\end{figure*}
%2028.oiioiiifit & 5321.oiioiiifit script (SM)

%%%%%%%%%%%%%%%%%%%%%%%%%%%%%%%%%%%%%%%%%%%%%%%%%%%%%%%%%%%%%%%%%%%%%%%%%%%%%%%%%%%
\begin{table*}
\footnotesize
\centering
\caption{Velocity measurements}
\begin{tabular}{lcccc|ccc}
 & & $FWHM_{NLR}$ & $FWHM_{out}$  & $V_S$  &  $\Delta V$  & $v_{max}$ & $w80$ \\ 
\cline{3-8}
& &\multicolumn{6}{c}{(km s$^{-1}$)}\\[0.5ex]\hline
\multicolumn{4}{l}{XID 5321}\\[0.5ex]% & & &  & &\\
\hline
$[OIII]$ & a & - & - & - & - & - & -\\ 
$[OII]$ & & - & - & - & $256 \pm 15$ & $950 \pm 70$ & $1040 \pm 55$\\ \cline{6-8} %3simm
$[OIII]$ & b &$ 310\pm15$ & $1320\pm10 $ & $350\pm10 $ & $500\pm20$ & $1730\pm 20$ & $1450\pm 25$\\
$[OII]$ & & - & - & - & $280 \pm 20 $ & $1050 \pm 25$ & $1060_{-20}^{+60}$\\ \cline{6-8}
$[OIII]$ & c & -  & $1650\pm50 $ & $330\pm20 $ &  $600\pm 200$ &$1950\pm 30$ & $2300\pm140$\\
$[OII]$ & & - & - & - & $225\pm65$ & $1350_{-60}^{+110}$ & $1230_{-100}^{+200}$\\ \cline{6-8}
\hline
\multicolumn{4}{l}{XID 2028}\\[0.5ex]% & & &  & &\\
\hline
$[OIII]$ &a & - & $1300\pm 45$ & $-385\pm15$& $-610_{-230}^{+60}$ & $-1530_{-310}^{+110}$ & $1230_{-50}^{+80}$\\ %2simm
$[OIII]$ &b & $510\pm13$ & $1310\pm65$ & $-340\pm30$ &  $-350\pm50$ &$-1350\pm50$ & $980\pm40$ \\ %1simm
$[OIII]$ & c & - & - & - & - & - & -\\
$[OII]$ & Keck & - & - & - & $-100\pm15$ & $810\pm20$ & $-900\pm100$\\ %2simm
\hline
\label{tvel}
\end{tabular}
\tablefoot{The values in the first three columns, corresponding to the simultaneous fit results, regard all the emission lines and not only the [OIII] emission. $FWHM_{NLR}$ and $FWHM_{out}$ refer to NLR and outflow components described in Section \ref{secsimultanei}.}
\end{table*}
%%%%%%%%%%%%%%%%%%%%%%%%%%%%%%%%%%%%%%%%%%%%%%%%%%%%%%%%%%%%%%%%%%%%%%%%%%%%%%%%%%%%

%%%%%%%%%%%%%%%%%%%%%%%%%%%%%%%%%%%%%%%%%%%
\section{ Additional evidences of outflow}

\subsection{[OII] emission lines}\label{secOII}

\citet{Zakamska2014} found that [OII]$\lambda\lambda$3726,3729 also show outflow signatures, in some cases consistent with extremely broad features as seen in [OIII]. 
At the redshift of our targets, the [OII] lines are redshifted in the 9200-9600 \AA\ range, sampled by the VIS arm of the X-shooter spectra. 

The rest frame wavelengths are inferred by the systemic redshifts obtained in the NIR spectrum (B14). A possible shift between the two different arms of X-shooter instrument could affect the rest frame wavelengths by $\sim$ 10 km s$^{-1}$ (see B14 Section 4.1). % (XID2028 Keck spectrum offset: 8 km s$^{-1}$;XID5321 X-shooter spectra offset: 47 km s$^{-1}$). 
The [OII] emission lines were fit with the non-parametric approach described previously, and the results are collected in Table ~\ref{tvel}. Given the lower S/N of [OII] with respect to the previously analyzed lines, and that the wavelength separation between the doublet lines is comparable with the resolution, in the fits we have assumed a 1:1 flux ratio  within the [OII] doublet and the emission lines were treated as a single Gaussian profile.% with 3727$\AA$ as the central rest frame wavelength.

\begin{figure}[!t]
\centering
\includegraphics[scale=0.3]{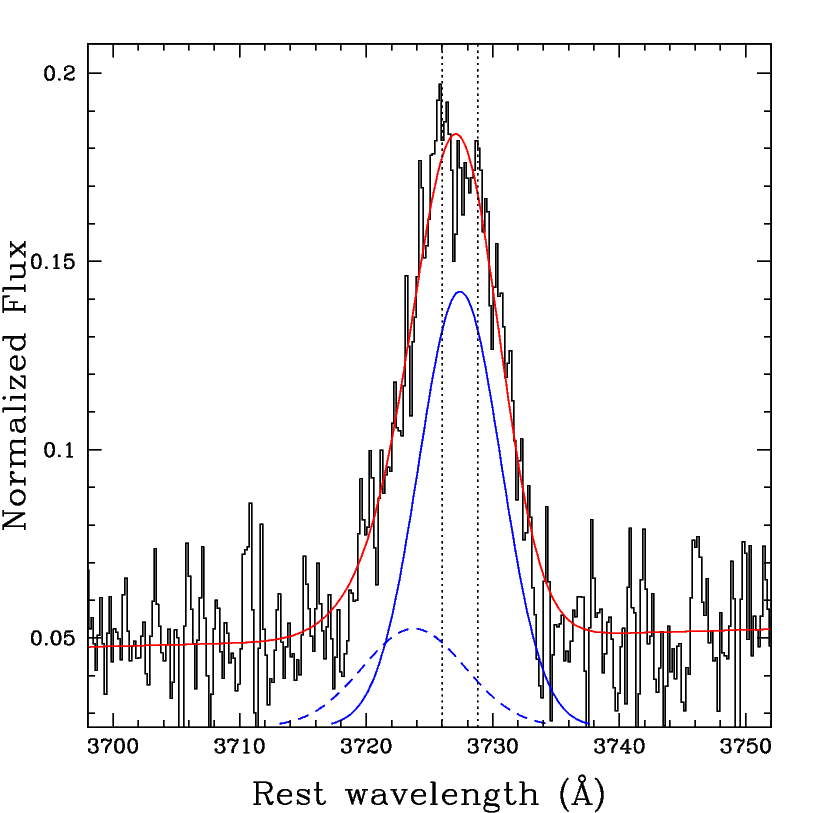}
%\captionsetup{font={small}}
\caption{Keck/DEIMOS spectrum around the [OII]$\lambda\lambda$3726,3729 region of XID2028. The spectrum is normalized so that the intensity of the strongest emission line (in the VIS+NIR wavelength range) of the nuclear aperture is unity. Superimposed on the spectrum are the best fit non-parametric analysis components (solid and dashed blue curves, with arbitrary normalization in order to ease the visualization). The red solid curves represent the sum of all components, including the power-law. Dotted lines mark the wavelengths of the [OII] emission lines.}
\label{fitOIIK}
\end{figure}

The [OII] emission lines of XID2028 are at the edge of the spectral range covered by the VIS arm, where the total transmission is very low. For this reason, we reanalyzed the available Keck/DEIMOS spectrum (presented in Fig. 13, B10). 
The Keck spectrum of XID2028, obtained with DEIMOS (in MOS mode) at Keck-II telescope (\citealt{Faber2003}) on 2008 January 8, with mildly lower resolution (2.5$\AA$; seeing $\sim$ 1''), was flux-recalibrated in order to renormalize it to the X-shooter spectra by multiplying the spectrum by a constant factor. %The rest frame wavelengths are inferred by the redshifts obtained in the X-shooter NIR spectrum. 
The best fit solutions are shown in Figure ~\ref{fitOIIK}: two Gaussian lines, one of them blueshifted with respect to the systemic velocity, are required to reproduce the shape of the emission line. As mentioned above, we have not tried to deblend the doublet, but a hint of the peaks could be in the vicinity of the dashed vertical lines.
The [OII] velocity offset of XID2028 is lower ($\sim$30\%) than the [OIII] one, but still statistically significant, and support the presence of outflowing ionizing gas also in this lower (but comparable to the [SII]) ionization phase.

\begin{figure}[!t]
\centering
\includegraphics[scale=0.3]{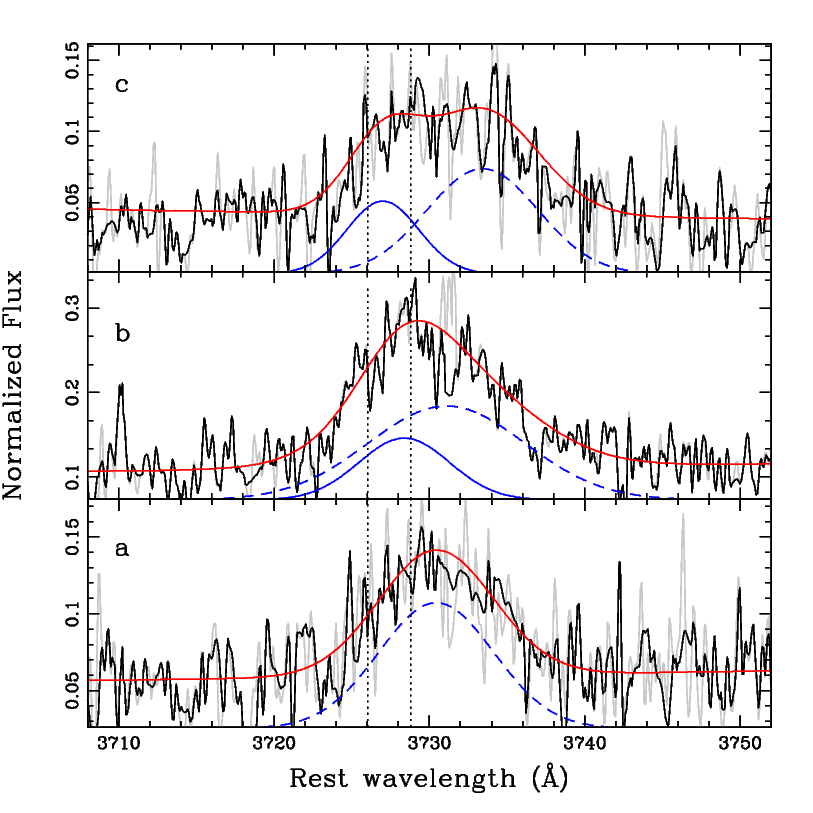}
%\captionsetup{font={small}}
\caption{X-shooter spectra around the [OII]$\lambda\lambda$3726,3729 region of XID5321. Grey and black lines indicate the original and the cleaned (as above) spectra. All the regions ($a, b, c$) are shown. See previous figure for description.}
\label{fitOIIX}
\end{figure}

As for the NIR, we extracted the XID5321 VIS spectra from the same three regions along the slit.
The [OII] emission lines of XID5321 are detected in all the three apertures (see Figure \ref{fitOIIX}), and therefore show a greater spatial extension with respect to the [OIII] emission line. 
%allowing us to fit the line profiles in all the three apertures.
%The two sources show the doublet nature of [OII]. 
The best fitting models are shown in Figure ~\ref{fitOIIX} and include two components for the nuclear ($b$) and off-nuclear ($c$) X-shooter spectra, and one component for the off-nuclear ($a$) spectrum. 
%The [OII]  intensity ratio of the doublet, like as those of the [SII], changes as function of the electron density of the environment. The low and high density limits ($0, +\infty$) correspond with the forbidden fine structure line ratio [OII] I(3729)/I(3726) of 0.35 and 1.5, respectively. In the fitting procedure we constrained the [OII] lines ratio to lie within this region.
%The fitting procedure has been implemented in order to add the ratios as free parameters. 
The wavelengths of the [OII] doublet are indicated with dashed vertical lines; here,  again, a hint of peaks is in the vicinity of the systemic [OII] lines. Nevertheless, we stress that the lines were fit with the non-parametric approach, and hence the individual Gaussian profiles shown in the figure (solid and dashed blue curves) do not have direct physical interpretation.
Even in this case, the [OII] velocity offsets of XID5321 are lower ($\sim$40\%) than the [OIII] one, but still significant. 

%FWHM$_N$ 101 km s$^{-1}$ (it is comparable to the narrower component found in the OIII emission line, 82km s$^{-1}$), FWHM$_B$=730 km s$^{-1}$. {\bf Fitted [OII] narrow lines are shifted of $\sim$ -48 km s$^{-1}$ compared to the systemic obtained in NIR spectra. This value is comparable to the resolution of the instrument in the VIS arm, $\sim$ 40 km s$^{-1}$.}    

 The comparison between the [OIII]$\lambda$5007 and [OII] doublets are shown in Figure ~\ref{profilivel}: [OII] velocity profiles are narrower than [OIII], but also shows almost the same asymmetries and shifts, in agreement with the results of \citet{Zakamska2014}.

Summarizing, %the fit to the XID2028 [OII] lines indicates the need for an additional  broad and blueshifted component to reproduce the line profile in the single aperture available from the DEIMOS spectrum.  
%The XID5321 [OII] emission lines show broad and shifted profiles in both lateral apertures, in contrast with the [OIII] emission line.
although at lower S/N, these emission lines confirm the presence of outflowing gas in both targets, and in one case (XID5321) with a greater spatial extension than in the [OIII] emission. 

%%%%%%%%%%%%%%%%%%%%%%%%%%%%%%%%%%%%%%%%%%%

\subsection{Absorption Lines}\label{secabsorp}

\begin{figure*}
\begin{minipage}[b]{9.0cm}
\centering
\includegraphics[width=7cm,height=7cm,angle=90]{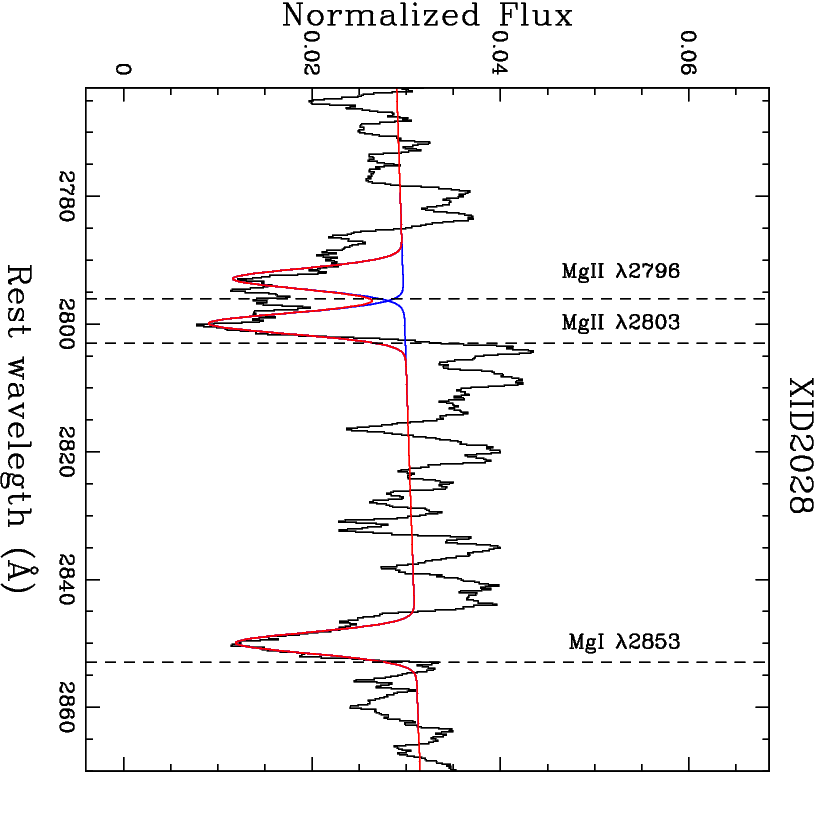}
\end{minipage}
\hspace{10mm}
\begin{minipage}[b]{9.0cm}
\centering
\includegraphics[width=7cm,height=7cm]{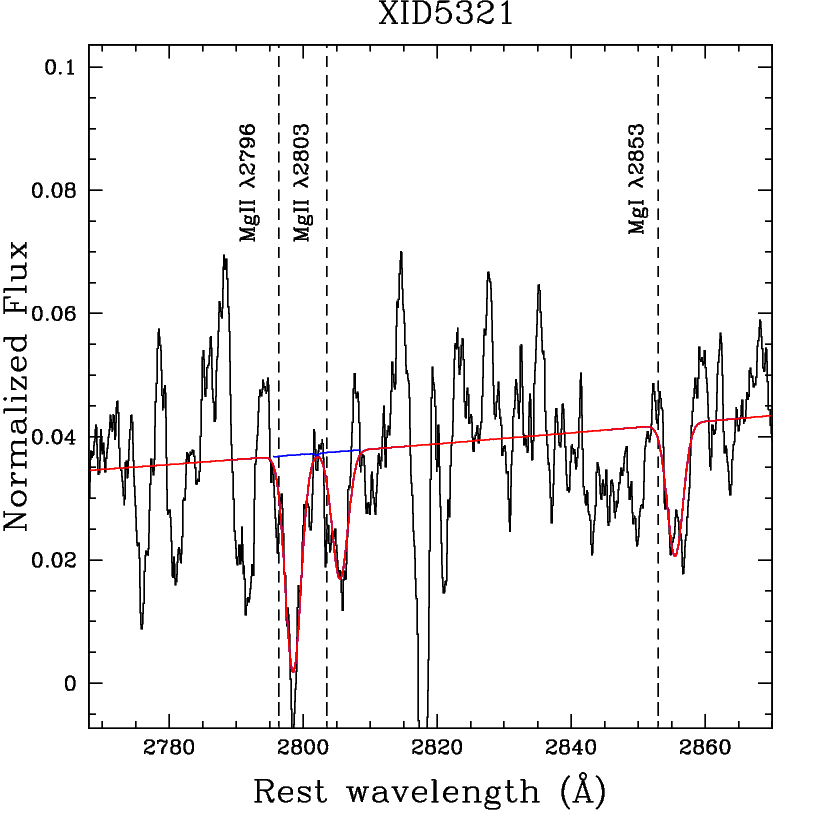}
\end{minipage}
\caption{XID2028 smoothed Keck/DEIMOS spectrum (left) and XID5321 smoothed X-shooter spectrum (right) with the magnesium absorption lines marked. The dashed lines in the spectra mark the rest-frame wavelengths of MgII and MgI, as determined from the systemic redshift. Superimposed on the spectra are the best fit components indicated as solid blue Gaussian curves. The red solid curves represent the sum of the Gaussian components. }
\label{assorbimenti}
\end{figure*}

Kpc-scale outflows are observed in both the neutral and ionized gas phases  \citep[e.g.,][]{Rupke2011,Rupke2013}. The sodium NaD$\lambda\lambda$5890,5896, magnesium MgII$\lambda\lambda$2796,2803 and MgI$\lambda$2853 absorption lines are detected in our spectra.
Figure ~\ref{assorbimenti} shows the Keck/DEIMOS spectrum of XID2028 (on the right) and the X-shooter NIR spectrum of XID5321 (on the left) with the magnesium absorption lines labeled. The Keck spectrum was preferred to the X-shooter one due to its higher S/N. %; the insets instead shows the velocity profile of the MgII and MgI lines. 
%Figure \ref{5321assorbimenti} shows the X-shooter NIR spectrum of XID5321 with the same absorption lines labeled. %As previosly, the insets show the velocity profile of the lines. 
Given the low S/N only the central aperture of XID5321 has been analyzed.

Each of the magnesium absorption lines of XID2028 were fit with one set of Gaussian profiles, as shown in Figure ~\ref{assorbimenti} (left). %{\bf In particular, it is the narrow MgI$\lambda2853$ profile that rule out a (more obvious, given the low S/N) 1 component fit: in fact, in that case, there would be a great residual in the bluer fraction of the MgII profile.} 
A velocity shift $V_S \approx$ -330 km s$^{-1}$ with respect to the systemic velocity was found.
%The corresponding velocity offsets is $V_S =$ -330 km s$^{-1}$. 
%The sodium absorption lines region is affected by numerous sky lines. 
% and therefore was fitted with one Gaussian component too.% constraining the relative velocity shift values obtained in the Mg absorption lines. Regions affected by sky features are excluded from the fit. 
%Given the possible shift between the VIS and NIR (Keck and X-shooter spectra), in the fitting procedure, the center of the bluer Gaussian profile was used as a free parameter, together with the amplitudes and the widths of the Gaussian profiles. 
%%%Figure ~\ref{2028assorbimenti} (right) shows the best fit solution of the sodium D region. 
The sodium profiles, although affected by numerous sky features, seem to be reproduced with the same velocity component of the magnesium profiles. %%%A shift  between the VIS and NIR (Keck and X-shooter spectra) can be estimated by comparing the wavelengths of the center of the magnesium and sodium Gaussian profiles with those of the systemic redshift and assuming the same velocity shift for both absorption lines. A NIR-VIS shift of $\sim$ 30 km s$^{-1}$, still lower or comparable to the resolutions, was found.

Each of the magnesium lines of XID5321 were fit with one Gaussian component too, as shown in Figure ~\ref{assorbimenti} (right). The corresponding velocity shift is $V_S \approx$ 260 km s$^{-1}$. Even for this source, the sodium profiles were reproduced with the same velocity component of the magnesium profiles. %Given the numerous sky lines in the region, the continuum cannot be fiducially estimated, and this fit allows us just to confirm the presence of a redshifted component in NaD as in Mg lines. %A shift VIS-NIR of $\sim$ 60 km s$^{-1}$ was found. %This shift, together with the difference in the FWHM of magnesium ($\sim$ 300 km s$^{-1}$) and sodium lines ($\sim$ 100 km s$^{-1}$),{\bf even if this could be reasonably due to a wrong estimate of the continuum in the vicinity of the sodium lines}, may suggest that we are observing two different regions with different kinematics. 
%However, both magnesium and sodium absorption lines are doubtless redshifted. 

We stress that for both sources, the sodium regions are strongly affected by sky features; the estimated continua and the Gaussian components did not allow us to study the properties of the absorbing gas, and are presented uniquely as proof of the constrained magnesium components, themselves characterized by low S/N. 

The estimated shifts VIS-NIR of a few tens of %$\approx$ 30 km s$^{-1}$ and 60 
km s$^{-1}$ between magnesium and sodium lines of XID2028 and XID5321 respectively, are greater than the accuracy of the wavelength calibration ($\sim$ 10 km s$^{-1}$; $\Delta z =$ 0.0004) checked with the position of known sky lines in both VIS and NIR spectra (see B14 Section 4.1). However, given that all our absorption lines suffer of low S/N, they could be fully explained by an imperfect determination of the center of the fitted Gaussian profiles and interpreted as residual error of the (greater) velocity shift $V_S$ values observed.

%Given that MgI has an ionization potential just 2.5 eV higher than the NaD absorption lines 
The XID2028 absorption lines can be ascribed to an outflowing neutral/low-ionization component on the line of sight, with roughly the same velocity shift of the emission lines. The XID5321 absorption lines show a velocity shift of the same magnitude of the emission lines too; possible interpretations are discussed in Section \ref{secneutral} and \ref{secsummary}.

\section{Line ratios}\label{seclineratio}

\subsection{Estimates of extinction}\label{secredd}
We used line ratios to estimate the reddening of the sources. 
This is crucial to derive the corrected [OIII] luminosity used later to estimate the outflow power. 
We used the Balmer decrements $I(H\alpha)/I(H\beta)$, as measured in our simultaneous fits described in Section 3.1,  where $I$ represents the line flux. These can be determined for each set of Gaussian component of the simultaneous fit. In particular, for the nuclear apertures, we used i) %the total fluxes of the Balmer emission line (summing the N, B, VB components resulting by the simultaneous fits) \\ 
%oppure: 1. 
the fluxes of the BLR components,
%Rose+13 non trova differenze tra i due metodi.
ii) the fluxes of the systemic NLR components, 
iii) the fluxes of the outflow components of the spectra. For the off-nuclear apertures, only one ratio can be estimated. 
When no H$\beta$ systemic or/and outflow component were detected, we used the standard deviation of the continuum regions (in the vicinity of the emission line) and the FWHM of that set of Gaussian profiles to derive a 3$\sigma$ upper limit on the flux. %and therefore a lower limit on the E$(B-V)$. 
Balmer decrements are reported in Table ~\ref{Balmer}.

\begin{table*}
\centering
\begin{minipage}[h]{15cm}
\footnotesize
\centering
\caption{Balmer decrements}

\begin{tabular}{l |cc | cc| cc}
% & VB & N & B \\ 
 & \multicolumn{2}{c|}{BLR} & \multicolumn{2}{c|}{NLR} &\multicolumn{2}{c}{outflow}\\[0.5ex] \cline{2-7}
 & $I(H\alpha)/I(H\beta)$ & $A_V$\footnote{Assuming Case B ratio of 3.1 and SMC extinction curve.} & $I(H\alpha)/I(H\beta)$ & $A_V$ &$I(H\alpha)/I(H\beta)$ & $A_V$\\
\hline
\multicolumn{4}{l}{XID 2028}\\[0.5ex]% & & &  & &\\
\hline
nuclear (b) & $ 4.4\pm0.3$& $0.9\pm0.4$ & $3.2\pm0.6$& $0.1\pm1.1$ & $6.2\pm 0.5$& $1.8\pm0.5$\\
off-nuclear (a) & -& -& -&-  & $>1.7$& -\\
\bottomrule
\multicolumn{4}{l}{XID 5321}\\[0.5ex]% & & &  & &\\
\hline
%\midrule
nuclear (b) & $6.4\pm0.5$ & $1.9\pm0.5$ & $>3.8$ & $>0.6$ &$>3.7$& $>0.5$\\
off-nuclear (c) & -  & - & - & - &$>1$& -\\
\bottomrule
\end{tabular}
\label{Balmer}
\end{minipage}
\end{table*}

%AV=1.97*log(balmerratio/3.1)*3.1
%A_5007 per smc=AV/3.1*3.2
The $I(H\alpha)/I(H\beta)$ ratio 3.1 (Case B, \citealt{Gaskell1984})  is mostly used to determine the amount of extinction for low-density gas, such as that of the NLR. The average value of BLR Balmer ratio of 3.06 obtained by \citet{Dong2008} for a large, homogeneous sample of $\sim$ 500 low-redshift type 1 AGN with minimal dust extinction effects, suggest that we can assume a ratio of 3.1 also for the BLR. We assumed the same ratio also for the outflow components.   %D. Ilic' Popovic' La Mura 2012 
Therefore, assuming Case B ratio of 3.1 and the SMC dust-reddening law, the relation between the emission line color excess and the Balmer decrements gives us, for XID2028, a V-band extinction $A_V$ in the range 0.1-1.8, where the first value is evaluated for the NLR component and 1.8 for the outflow component. For XID5321, we obtain an extinction $A_V$ in the range 0.5-1.9, where the first value is a lower limit evaluated for the outflow component, and 1.9 is obtained for the BLR region. Ratios from the off-nuclear apertures are smaller than 3.1, and no $A_V$ can be derived. Lower limits on the extinction follow from the use of upper limits on the flux of undetected H$\beta$ emission lines. 
In the following we will use the outflow component Balmer ratios to determine the extinction of the [OIII] outflowing flux. In fact, those obtained in the SED fitting procedure are related to the continuum (BBB) flux that, being affected by a degeneracy between host and AGN accretion disk emission, are significantly more uncertain.

\subsection{Rest-frame optical AGN-SF diagnostics}\label{secbpt}

\begin{figure}
\centering
\includegraphics[scale=0.3]{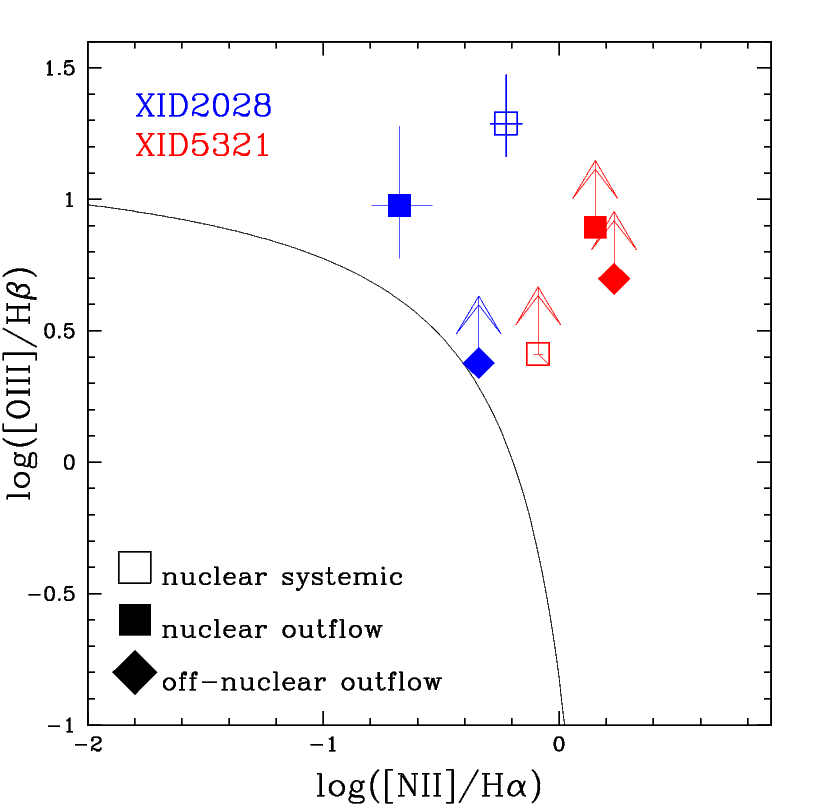}
%\captionsetup{font={small}}
\caption{Standard diagnostic diagram showing the classification scheme by \citet{Kewley2013}. The line drawn in the diagram corresponds to the theoretical redshift-dependent curve used to separate purely SF galaxies from galaxies containing AGN (\citealt{Kewley2013}). Empty and filled squares correspond to the systemic and outflow components of the nuclear aperture; filled diamonds correspond to the outflow components of the off-nuclear apertures. Blue and red points represents XID2028 and XID5321 respectively. Upper arrows represent lower limits, due to undetected H$\beta$ emission lines.}
\label{bpt}
\end{figure}

%Our simultaneous fits with [OIII] and H$\alpha$ regions allow us to investigate the BPT diagnostics. 
We used  the optical diagnostic Baldwin-Phillips-Terlevich diagram (BPT; \citealt{Baldwin1981}) as a tool to investigate the nature of the ionizing sources. Figure ~\ref{bpt} shows the BPT with the results from our spectroscopic analysis. The line drawn in the diagram corresponds to the theoretical redshift-dependent curve used to separate purely SF galaxies from galaxies containing AGN at z=1.5 (Eq. 1 of \citealt{Kewley2013}). %Empty and filled squares correspond to the systemic and outflow components of the nuclear apertures; filled diamonds correspond to the outflow components of the off-nuclear apertures. Blue and red points represent XID2028 and XID5321, respectively. Upper arrows in the figure represent lower limits, due to non-detected H$\beta$ emission lines. 
For both sources, all the systemic and the outflow components observed in
the different apertures %for both sources
%the points in the figure 
are consistent with an AGN classification.

%-Contribution from Star Formation?
%(Lal \& Ho 2010):
We tested other possible line ratio diagnostics \citep[e.g.,][]{Kewley2006,Ho2005}, involving [OIII]/[OII] and [OI]$\lambda$6300/H$\alpha$ (detected only in the nuclear spectrum of XID5321). 
%Using the OI/H$\alpha$ ratio it is possible to put these value in a different BPT diagram ([OIII]/[OII]-[OI]/H$\alpha$; \citealt{Kewley2006}, see Sec. 3). Unfortunately, the [OI]$\lambda$6300 is detected only in the central spectrum of XID5321; the ratios are located in the AGN region of the diagram, for both the narrow and broad components. % (log(OIII/OII)$_N$=0.2, log(OI/H$\alpha$)$_N$=-0.9; log(OIII/OII)$_B$=0.9, log(OI/H$\alpha$)$_B$=-0.5, see Fig. 5 of Kewley et al. 2006).\\
%Ratios $[OIII]\lambda5007/[OII]\lambda3727,29$ within the range 5-10 are expected from AGN photoionization regions (Ho 2005). 
%The ratios %of log([OIII]/[OII])$_N$=0.8, log([OIII]/[OII])$_B$=0.7 
%obtained from the narrow and broad components of the oxygen lines of XID2028, [OIII]/[OII], are in the range expected from AGN photoionization (Ho 2005). Therefore, despite the undetected [OI]$\lambda$6300 of XID2028, that did not permit to localize its components in this second BPT, we may assume that 
%the [OII] as well as [OIII] emission lines are mainly produced from AGN photoionization regions for both targets.
%A third BPT diagram involving the [OIII]/H$\beta$-[SII]/H$\alpha$ ratios is not taken into account given the weakness and strong degeneracy in the fitting procedure below the sulfur emission lines. 
All of them agree with an AGN photoionization origin. We can consequently conclude that, even if present, a star forming origin would play a minor role (similar conclusions have been reached in \citet{Zaurin2013}). %, but see also discussion on shocks there).

\section{Quantifying the outflow energetics}
Quantifying outflow energetics require estimating the kinetic power ($P_K$) and mass-flow rate ($\dot M$) of the outflows. These quantities can be computed when we are able to estimate the temperature, the electron density $n_e$, the involved emitting luminosity, the spatial scale, i.e. the distance R of the outflowing material from the central source, the outflow velocity, and to infer the geometry distribution. In addition, all the gas components should be probed.

\subsection{Ionized component}\label{secioniz}

Following the arguments presented in \citet{CanoDiaz2012}, the kinetic power associated with the ionized component of the outflow can be given by:

\begin{equation}\label{canodiaz}
P_K^{ion}=5.17\cdot 10^{43} \frac{CL_{44}([OIII])v_{0,3}^3}{n_{e3} R_{kpc} 10^{\left [ O/H \right ]}} erg\ s^{-1},
\end{equation}

where $L_{44}([OIII])$ is the [OIII] luminosity associated to the outflow component in units of 10$^{44}$ erg s$^{-1}$ and corrected for the extinction, $n_{e3}$ is the electron density in units of 1000 cm$^{-3}$, $v_{0,3}$ is the outflow velocity $v_0$ in unit of 1000 km s$^{-1}$, $C$ is the condensation factor ($\approx$ 1), 10$^{[O/H]}$ is the metallicity in solar units, $R_{kpc}$ is the radius of the outflowing region in units of kpc. 

Our slit-resolved spectroscopy analysis has shown that significant signal in the [OIII] shifted components can be detected out to a projected distance of R $\approx$ 11 kpc (see Figure ~\ref{aperture2D}). Therefore we assume the outflow extending out to that radius from the central black hole for both sources (see also \citealt{Cresci2014}). 

Estimates of the electron density of the outflowing gas from the flux ratios of the outflow components in the nuclear apertures of the [SII] doublets can be obtained (\citealt{Osterbrock1989}): $n_e\approx 2000-3000$ cm$^{-3}$ (B14). Although these values are higher than other values routinely used in Eq. ~\ref{canodiaz} \citep[e.g.,][]{Liu2013,Harrison2014}, they are still comparable with other assumptions adopted in low-z AGN/ULIRGs systems (e.g., \citet{Zaurin2013,Villar2014} and references therein). However, these $n_e$ estimates are somewhat ambiguous, given the complexity of the spectral profiles. Moreover, when systemic components are also present, the fitting procedure may produce strongly degenerate results. For these reasons, we use the value obtained from the fit to the [SII] doublet in the off-nuclear aperture ($c$) of XID5321, $n_e=120$ cm$^{-3}$. This electron density, close to the commonly used in similar studies, is adopted for both targets (unfortunately, the off-nuclear ($a$) aperture of XID2028 did not permit a similar estimate).

Estimates of the metallicity, using the indicators $N2=[NII]\lambda6583/H\alpha$ (\citealt{Pettini2004}) and $R23 = ([OII]\lambda\lambda3726,3728 + [OIII]\lambda\lambda4959,5007)/H\beta$ \citep[e.g.,][]{Pilyugin2001,Yin2007}, based on the assumption of a stellar ionizing radiation field, are not useful because a large contribution from the AGN is present for both sources. % (see, e.g., Kewley \& Ellison 2008). 
For this reason, we assume a solar metallicity. In fact, even assuming that the outflows reside in metal-rich regions with $9.0<12+log(O/H)<9.3$ (\citealt{Kewley2013,Du2014}), the metallicity would be at most a factor of 2 greater. This is negligible with respect to other sources of uncertainties (see below). 

From the extinctions obtained previously from Balmer decrements of outflow components, we obtain values of $A\_5007\AA\approx1.8$ and $0.5$ (lower limit), corresponding to correction factors of $\approx$ 6 and 2 for the [OIII] luminosities of XID2028 and XID5321, respectively.
%Assuming the reddening law of Cardelli et al. 1989, there is ongly an 8\% difference in the extinction of Hb and OIII

Finally, following \citet{CanoDiaz2012} and \citet{Cresci2014}, we assume that the maximum velocity observed $v_{max}$ is indicative of the outflowing velocity of the gas, while lower velocities are due to projection effects. %$v_{max }$ is the maximum projected velocity measured, and regards only the 2\% of the gas producing the emission lines. 

Using the values previously indicated and the measured $v_{max}$, we obtain

\begin{equation*}
P_k^{ion}(2028)\approx4\cdot 10^{43} erg/s,
\end{equation*}
\begin{equation*}
P_k^{ion}(5321)\approx6\cdot 10^{43} erg/s, 
\end{equation*}

adding the contributions determined in the apertures $(a)+(b)$ and $(b)+(c)$ for XID2028 and XID5321. Namely, for each object, we compute the relative kinetic power adding the corresponding flux measured in the outflow components, and using the maximum $v_{max}$ value (see Table \ref{tvel}). %\footnote{These values are a factor 10-60 greater than the corresponding values quoted in B14. The main difference comes from the different electron density adopted (a factor of 8 difference) and the correction for the extinction (see discussion in Sec. 6 of B14).}.
Consequently,
using the same assumptions as above and Eq. B8 of \citet{CanoDiaz2012}, we obtain the corresponding outflow mass rates $\dot M_{out}^{ion}(2028)\approx 55\ M_{\odot}\ yr^{-1}$ and $\dot M_{out}^{ion}(5321)\approx 50\ M_{\odot}\ yr^{-1}$. %

Equation~\ref{canodiaz} and Eq. B.8 of \citet{CanoDiaz2012} assume a simplified model where the wind occurs in a conical region composed of ionized clouds uniformly distributed, under the assumption that most of the oxygen is in the O$^{+2}$ form. Because of this assumption, these values are lower limits for the kinetic power and mass rate of the ionized component. 

A confirmation that the estimates from [OIII] are lower limits, comes from the outflow mass ratio %of $\approx 250 M_{\odot}\ yr^{-1}$ 
computed comparing the [OIII] ionized mass %from Eq. B.5 of \citet{CanoDiaz2012}
with the H$\beta$ mass estimate, %from Eq. 14 of \citet{Liu2013}
using the measured electron density and assuming an electron temperature $T_e=2\cdot10^4$ K (\citealt{Osterbrock1989,CanoDiaz2012}): $M_{[OIII]}/ M_{H\beta}=0.008\ L_{[OIII]}/ L_{H\beta}$. %, using the observed H$\beta$ flux evaluated in the nuclear region, and the same kinematics and extension of the [OIII] emission lines in the same region. 
%The corresponding outflow mass rate ratio, %estimated by the continuum fluid equation, $\dot M=3 M v_0/R$ (see \citet{Cresci2014} for details), 
If we assume the same kinematic and extension for H$\beta$ and oxygen emission lines, the corresponding outflow mass rate ratio preserves the same dependency: $\dot M_{[OIII]}/ \dot M_{H\beta}=0.008\ L_{[OIII]}/ L_{H\beta}$. The same applies to the kinetic power ratio.\\
XID2028 measured luminosity ratio $L_{[OIII]}/ L_{H\beta}\approx$ 10 in the nuclear region (i.e., where outflowing H$\beta$ is detected), implies that we are underestimating the ionized mass rate by a factor of $\approx 10$.
XID5321 upper limit on the flux of undetected outflowing H$\beta$ emission line in the nuclear region (see Section \ref{secredd}) implies at most a factor of $\approx 20$ larger ionized mass rate. 
As a conservative order of magnitude estimates of the ionized kinetic power and mass rate of the targets we will refer, in the following, to the values previously evaluated, multiplied by 10, in order to account for the  $L_{[OIII]}/ L_{H\beta}$ ratios. Therefore, our inferred mass rates are $\dot M_{out}^{ion}(2028)\approx 550\ M_{\odot}\ yr^{-1}$ and $\dot M_{out}^{ion}(5321)\approx 500\ M_{\odot}\ yr^{-1}$, and the corresponding kinetic powers are $P_k^{ion}(2028)\approx4\cdot 10^{44} erg/s$ and $P_k^{ion}(5321)\approx6\cdot 10^{44} erg/s$.
For XID2028, these results are consistent with those obtained from SINFONI near infrared integral field spectroscopy observations: \citet{Cresci2014} estimate from the H$\beta$ flux an outflow mass rate $>300\ M_{\odot}\ yr^{-1}$, without extinction correction.

%Indeed, when the assumptions on metallicity and on the O$^{+2}$ can be removed, i.e. using the H$\beta$ emission line obtained from SINFONI near infrared integral field spectroscopy observations available for XID2028 (092.A.0144, PI: Mainieri), \citet{Cresci2014} estimate an outflow mass rate of $300\ M_{\odot}\ yr^{-1}$ for the ionized component of XID2028. The $H\beta$ flux used to estimate the mass rate has been measured on the revealed outflow region (see their Fig. 1). Although a possible contribution from the unperturbed ionized NLR gas in the outflow region could still be present, their H$\beta$ flux was not corrected for the extinction; therefore our XID2028 outflow mass rate is certainly a lower limit. 

%In order to quantify the uncertainties induced by the lack of measurement of the electron temperature,
We also derive estimates of the mass rate in more extreme conditions, assuming an electron temperature $T_e=5\cdot10^4$ K, a rescaled electron density ($n_e=10^2\cdot T^{0.5}\cdot(r-1.49)/(5.62-12.8r)$, with $r=I([SII]\lambda6717)/I([SII]\lambda6731)$; \citealt{Acker1995}), and $\dot M_{[OIII]}/ \dot M_{H\beta}$ relation. % with [OIII] and H$\beta$ emissivities related to these conditions (\citealt{}), 
These conditions are chosen in order to obtain the lowest values of mass rate and $\dot M_{[OIII]}/ \dot M_{H\beta}$ correction factor. We obtain $\dot M_{out}^{ion}\approx100\ M_{\odot}\ yr^{-1}$, for both the sources. This value, therefore, could be taken as a very conservative lower limit of the total ionized outflow mass rate.

\subsubsection{Kinetic power $v_0$ estimators}

%°°°°Velocities - Kinetic power estimators°°°°% 
%On similar arguments, we also note that estimates of 
Usually, kinetic powers are determined taking as $v_0$ the velocity offset (defined as the offset between the Gaussian components of the fitting procedure as well as the non-parametric one, i.e. as $\Delta V$ as well as $V_S$; see, e.g., \citealt{Harrison2012} and references therein). However, these are projected velocities and therefore may be not representative of the true outflow velocities. \citet{Harrison2012}% proposed $v_0=FWHM/2$:
, found that $FWHM_{out}/V_S\approx$2 for two of their sources showing clear BLR components and, assuming that the outflow could be primarily oriented towards us for this kind of sources, suggest that FWHM$_{out}$/2 may be an adequate approximation of $v_0$ for other sources. However, we did not find the same ratios (the same conclusions are found for the non-parametric $w80$/$\Delta V$ ratios), suggesting a different geometry.

We also note that estimates of kinetic power through $w80$ value could be similarly arguable: our $w80$ values are lower in the nuclear region than in the off-nuclear ones.
This has possibly significant implications: $w80$ may be not reliable when there is a unique extracted spectrum instead of slit-resolved spectroscopy on different apertures, because the line flux of the systemic component in the core is dominating the emission. At larger radii this should not apply, because the NLR flux is almost suppressed. Therefore, $w80$ values greater in the external regions may be not interpreted as accelerating outflows.

We define the line width $w40$, as the velocity width $v50-v10$ that contains 40\% of the emission flux of the line profile of XID2028, with blueshifted outflow, and as the width $v90-v50$ of the line profile of XID5321, with redshifted outflow signatures (\citealt{Cresci2014}).
These estimators, regarding only the emission of more certain outflowing gas origin, can be useful to quantify possible outflow velocity variations between the different regions. XID2028 shows a roughly constant value ($w40=580\pm 30$ and $610\pm 40$ km s$^{-1}$ for the off-nuclear ($a$) and nuclear ($b$) regions respectively), and this may indicate that the outflow is not accelerating. %, as also confirmed by the statistically consistent w80 values in Table.
This is not the case for XID5321, where $w40$ shows a significant variation between the two regions ($w40=1730\pm20$ and $2380\pm200$ km s$^{-1}$ for the nuclear ($b$) and off-nuclear ($c$) apertures), mostly attributable to $v90$ changes. 
In this source, therefore, an accelerating outflow could be present. 

All these arguments suggest that the kinetic powers evaluated both in this study and in others using similar velocity estimators, should be considered only order of magnitude estimates. To give a better feeling of the uncertainties, for these two sources, $v_{max}$ and $w80$ are in reasonable agreement (within $10-30\%$), but $\Delta V \sim 0.3 v_{max}$ implies a variation of a factor of $\sim$ 40 in kinetic power estimates. Improved observational constraints are needed to test kinematic models of these outflows, and allow better estimation of kinetic energy and mass rates.

\subsection{Neutral component}\label{secneutral}

In the optical thin approach, from the ratio of the Equivalent Width (EW) for any two transitions of the same ion it is possible to estimate the ion integrated column density. %: the EW scales as $Nf\lambda_0$, where $f$ is the transition oscillator strength, $\lambda_0$ is the wavelength of the transition in Angstrom, $N$ is the density of ion. 
Namely, the EW doublet ratio can assume values in the range 2-1, corresponding to optical depth in the range $0,+\infty$ (\citealt{Bechtold2003}). 

The measured EW ratio of MgII$\lambda\lambda$2796,2803 of XID5321 is 1.7, therefore the magnesium column density can be inferred from the EW of MgII$\lambda$2796 (see \citealt{Bechtold2003} for more details). Following the arguments presented in \citet{Bordoloi2013}, i.e. assuming no ionization correction ($N_{Mg}=N_{MgII}$), no correction for saturation, and a solar abundance of Mg %($log(Mg/H)=-4.42$) 
with a factor of -1.3 dex of Mg depletion onto dust, we obtain a hydrogen column density $N_H(5321)=3\cdot 10^{19}$ cm$^{-2}$.

The measured EW of the magnesium lines of XID2028 are instead consistent with a doublet ratio of $\sim1$, corresponding to a high optical depth. In this case, the absorption lines would be approaching the flat part of the curve of growth and their column density cannot be inferred from their EW. We estimate an apparent optical depth and a conservative lower limit of the hydrogen column density,  $N_H(2028)>9\cdot$10$^{19}$cm$^{-2}$, following the method used by \citet{Bordoloi2013} and using the same assumptions as above.

For both sources, we have used the EW of low ionization MgII to estimate lower limit of neutral hydrogen column densities, instead of NaD because of the difficulties inherent the sodium fits (see Section \ref{secabsorp}). This corresponds to assume these absorption lines as in a unique absorbing system. The assumption is justified by the fact that all the absorption lines seem to have same kinematics. Moreover, given that they have roughly the same velocity shift of the outflowing ionized gas, we assume that this absorbing system has the same kinematic and extension of the ionized one, i.e. that neutral and ionized components are closely connected (see also discussion below). %If this was the case, the redshifted absorption lines of XID5321 could be explained with outflowing absorbers on the line of sight, illuminated by the light of the host galaxy behind (see Figure \ref{cartoon5321}; Section \ref{sec5321outflow}).

Corresponding neutral mass outflow rates can be estimated from Eqs. 7-8 of \citet{Weiner2009}, 
\begin{equation}\label{MHout}
\dot M_{out}^H\simeq 7\cdot \left ( \frac{N_H}{10^{20}cm^{-2}}\right ) \left (\frac{R}{5kpc} \right ) \left ( \frac{v_0}{300 km s^{-1}} \right )\ M_{\odot}\ yr^{-1}, 
\end{equation}

assuming a flow that extends from radius 0 to R. %For both sources, the magnesium velocity shifts ($\left | V_S(Mg) \right |\approx$ 300 km s$^{-1}$) are comparable, within the errors, to the oxygen values ($\left |V_S([OIII]) \right |\approx$ 350 km s$^{-1}$) of the central ($b$) apertures. This could suggest that the outflowing ionized and the neutral components are closely connected in these two sources. If this were the case, the redshifted absorption lines of XID5321 could be explained with outflowing absorbers on the line of sight, illuminated by the light of the host galaxy behind (see Section ??; Figure ??). 

For the assumptions discussed above, %If we assume that the neutral component has the same kinematics and extension of the ionized one%(given that some of the ionized outflowing gas becomes again neutral)
using the maximum $v_{max}$ value and the maximum extension observed for the [OIII] emission lines, we obtain mass outflow rate $\dot M_{out}^H(2028)>80\ M_{\odot}\ yr^{-1}$ and $\dot M_{out}^H(5321)>35\ M_{\odot}\ yr^{-1}$. Given the adopted assumptions, these results represent lower limits. Adding the previous results (ionized items), we obtain lower limits of the total ionized and neutral mass outflow rate:  $\dot M_{out}^{tot}(2028)>630\ M_{\odot}\ yr^{-1}$ and $\dot M_{out}^{tot}(5321)>535\ M_{\odot}\ yr^{-1}$. %an estimate of the molecular component involved is still missing).

\section{Discussion and conclusions}\label{secsummary}

We analyzed in great detail the optical, NIR and X--ray spectra of two luminous, obscured QSOs, selected from the XMM-COSMOS survey to be in the ``blow-out'' phase of galaxy-AGN co-evolution. Although not as extreme as other very luminous QSOs selected from all-sky surveys and thought to be caught in this crucial feedback phase (with L$_{\rm bol}>10^{47-48}$ erg s$^{-1}$, e.g. the WISE-selected quasars in \citealt{Weedman2012,Eisenhardt2012}, or the UKIDSS-selected objects in \citealt{Banerji2012,Banerji2014}), these two sources are among the most luminous objects at z$\sim1.5$ (L$_{\rm bol}>10^{46}$ erg s$^{-1}$) of the entire X--ray selected sample in the much smaller area covered by the COSMOS survey, thus more representative of the entire luminous population. The availability of a complete and deep multiwavelength coverage within the COSMOS field enables a full characterization of the host galaxy and accretion rates properties of these two extreme sources which is not possible, at similar depths, in other targets given the limited follow-up available.
%Although detailed multiwavelength investigations are still scarce given the limited follow-up available, there are indications that 

XID2028 and XID5321 are  ``X--ray loud'', as per their selection, based on optically dim counterparts associated to bright X--ray and MIR emissions. In addition, when compared with the average SED of Type 2 AGN with similar bolometric luminosity, the X--ray flux of XID2028 and XID5321 is less absorbed than average (the X--ray slope in Figure \ref{sed} is flatter than the \citet{Lusso2011} X--ray SED).  Indeed, thanks to the high quality X--ray spectra ($>1500$ counts) we were able to robustly constrain the column density in the range  N$_{\rm H}\sim6-7\times10^{21}$ cm$^{-2}$.

%Our targets, instead, have a much lower SFR (a factor of 10 less).  Moreover, 
Despite the relatively high SFR observed (SFR$\sim250$ M$_{\odot}$ yr$^{-1}$), the high X--ray luminosity coupled with the low level of column density observed in our two sources,
% N$_{\rm H}\sim6-7\times10^{21}$ cm$^{-2}$, constrained from the very high quality X--ray spectra ($>1500$ counts), 
argues against an obscured accretion phase (e.g., in evolutionary models when the SFR is at its peak, it should be associated with high nuclear and host obscuration; see, e.g., \citealt{Hopkins2008}). The overall properties extracted from the SEDs and the X--ray spectra, instead, definitely point towards a scenario in which  the AGN is caught in the subsequent phase where most of the BH mass has been already fully assembled. %and the X--ray emission is reaching its maximum (see \citealt{Hopkins2008}). 
This is witnessed by the Broad components observed in the H$\alpha$ region, corresponding to BH masses M$_{BH}\sim10^{9-10}$ M$\odot$ (\citealt{Bongiorno2014}). The fairly high accretion rates observed in our two targets ($L_{bol}/L_{Edd}\sim0.01-0.05$) are also in agreement with a scenario in which our targets are caught in the very short blow-out phase (see Fig. 2 of \citealt{Hopkins2005}), between the obscured and unobscured peaks of bolometric luminosity and accretion rate.
%As a side note, we also stress that the X--ray loudness implies that a lower bolometric correction should to be adopted (k$_{\rm {bol}}\sim4-10$) to go from the X--ray to the bolometric luminosity when compared to the average k$_{\rm bol}\sim12-15$ for objects at similar X--ray luminosity (\citealt{Lusso2012}). This is important when bolometric luminosities have to be derived by single band detections, in order to have an estimate of the AGN energetic input.

 The X-shooter (VIS and NIR) and Keck/DEIMOS spectra allowed us to simultaneously sample different gas components and phases, via spatially resolved spectroscopy of the emission lines complexes (H$\beta+$[OIII] and H$\alpha+$[NII]+[SII] in the NIR spectra, [OII] in the VIS and DEIMOS spectra). 
%and the study of the absorption systems (MgI, MgII, and NaD).
In particular, with respect to the analysis presented in B14, in where we discuss only the spectra of these two sources extracted in the nuclear regions, here we investigated the kinematic properties of the gas observed at projected  distances of 6-12 kpc. In addition, we studied the absorption properties of the MgI, MgII and NaD systems and we could put better constraints on the properties of the outflows, including their neutral components.

%°°°°Extension°°°°% 
 First of all, we found clear evidence for the presence of outflow components in all the emission 
%and absorption lines (!!!absorption lines only in central apertures!!!) 
lines investigated in one of the two off-nuclear apertures we extracted for each source (see Section \ref{secoutflowprop} and Figures \ref{aperture2D}, \ref{profilivel} and \ref{fitsimultanei}). In the hypothesis that these components are tracing outflowing winds, we were therefore able to demonstrate that the outflow has propagated well within the host galaxy (up to 10-12 kpc scale; see Section \ref{secextension}). 
The extension of the outflow region of XID2028 %, up to 10-12 kpc scale (see Section \ref{secextension}), 
has been confirmed by SINFONI J-band data, where the outflow has been unambiguously resolved and its impact on the host galaxy has been also extensively discussed (\citealt{Cresci2014}).
%These results have been confirmed for XID2028 by SINFONI J-band data, where the outflow has been unambiguously resolved and its impact on the host galaxy has been also extensively discussed (\citealt{Cresci2014}, submitted).

%°°°°Fit Methods°°°°% 
In order to constrain at best the velocity of the outflows, we performed both non-parametric analysis and simultaneous fit of the emission lines. The non-parametric measurements of the emission lines removed the degeneracy between the systemic and outflow components, allowing better constraints on the velocity properties (Section \ref{secnonparametric}), while the simultaneous fitting allowed us to quantify the relative outflow luminosities at different radii and to disentangle the contribution of the outflow fluxes from those of the NLR in the central apertures (Section \ref{secsimultanei}).%: the profiles of the central apertures (red curves in the figures \ref{profilivel} and \ref{fitsimultanei}) are broad and asymmetric; those in the lateral apertures (orange curves in the same figures) also appear broad and even more shifted with respect to the systemic.

With a solid handling of the most important observational constraints (velocity and spatial scale) needed to derive the outflow kinetic powers and mass outflow rates, and with the additional information we could derive from the absorption systems, we could draw conclusions on their associated energetics and mass outflow rates (see Section \ref{secneutral}). 
The kinetic power fiducial values obtained, being greater ($\gtrsim3$) than those presented in B14, can only be attributed to AGN-driven outflows and definitely rule out a SF origin for the outflow. %Indeed, even assuming a 100\% coupling of the kinetic power associated to the stellar winds (given the SFR$\sim250$ M$_{\odot}$ yr$^{-1}$; see Fig. 9 of B14) into energy needed to drive the outflows ($\dot{E}_{SF}\sim1.7\times10^{44}$ erg s$^{-1}$, following Veilleux et al. 2005), this would not be enough to obtain the total inferred kinetic powers ($>10^{45}$ erg s$^{-1}$). Instead, the AGN luminosities could be enough to sustain these: the values of $<P>$ are $\approx$ 10-20\% of the AGN bolometric luminosities, in reasonable agreement with the prediction by AGN feedback models given the large uncertainties involved (e.g., King 2005). 
In fact, an AGN origin of the outflow, with kinetic powers of $\approx$ 4\% of the AGN bolometric luminosity, is in agreement with the prediction by AGN feedback models (e.g., \citealt{King2005}), and is favored compared to the very high coupling needed between SF and the wind (see the detailed discussion of B14, Sec. 6.1, and references therein).
An AGN origin for the observed outflowing winds is also supported by our BPT analysis: all the outflow components observed in the different apertures for both sources lie in the region of AGN photo-ionization (see Figure \ref{bpt} and Section \ref{secbpt}).

Moreover, the momentum fluxes, defined as $\dot P=\dot M_{out}v_0$, inferred for these outflows are $\dot P(2028) \approx 6.1 \cdot 10^{36}$ dyne and $\dot P(5321)\approx 6.6 \cdot 10^{36}$ dyne. These values are in excess of $\approx 10$ times the radiative momentum flux from the central black hole, $\dot P_{rad}= L_{bol}/c$. This excess is consistent with the ``momentum boost'', $\dot P/\dot P_{rad}$, observed in local ULIRGs dominated by AGN and in luminous QSO (\citealt{Rupke2011,Sturm2011,Faucher2012}), and required to reproduce the normalization of the relation among BH mass and stellar velocity dispersion, $M_{BH}-\sigma$, in numerical simulations (\citealt{DeBuhr2012}).

Finally, we obtain the dynamical time $t_d\approx R/v_0$ of the outflow: $t_d(2028) \approx 8 Myr$ and $t_d(5321)\approx 5 Myr$. The dynamical time scales of these outflows are much longer than the recombination time scales of the [OIII] lines (a few tens of $yr$; e.g., \citealt{Osterbrock1989}), as also noted by other authors \citep[e.g.,][]{Cresci2014,Greene2012}. 
 %Otherwise, we would expect to observe negligible [OIII] flux. %However, the current data do not allow to discriminate between continuous and intermittent AGN fueling over the entire dynamical time period. 
Given the spatial extension of the [OIII] emission, we can prove the presence of AGN emission over a period of $t_e \sim 3\cdot 10^4 yr$. This time estimation comes from the size of the outflow divided by the velocity of light $c$. In fact, for these two sources, given that the AGN is the main source of O$^{+2}$ ionization, black hole accretion and consequent AGN emission are required. However, we can not discriminate between countinuous and intermittent AGN emission over the entire dynamical time period, if intermittent AGN fueling takes place on time scales much shorter than $t_e$, or on larger time scales.
Moreover, even the outflow could be continuously or intermittently refilled. %The current data do not allow to discriminate between these cases: h
Higher angular resolution data are required  to infer the distributions of the ejected material and to obtain more robust estimates of the outflow properties.

% or if it takes plase on time scales larger than that time scale could not exclude an intermittently fueled AGN over the entire dynamical time period, as also we could not discriminate between different distributions of the clouds ejected; higher angular resolution data are required \citep[for more details see e.g.,][]{Cicone2014}.

Despite a remarkable set of common properties in the two targets (similar host galaxies and accretion properties, similar broadening of the lines and spatial extension), XID2028 and XID5321 present one striking difference: the outflow component is {\it blueshifted} in XID2028 (as observed in basically all the objects with broad [OIII] reported in the literature) while in XID5321 the outflow component is {\it redshifted}. 
Below we will therefore summarize separately the properties of the two systems we were able to infer from all the available data. 

\subsection{The blueshifted outflow in XID2028}

%°°°°Absorption°°°°% 
% Similarly, our spectral analysis revealed blueshifted sodium (NaD$\lambda\lambda$5890,5896) and magnesium (Mg$\lambda\lambda$2796,2803 and MgI$\lambda$2853) absorption lines in XID2028% and XID5321, respectively. 
%For both sources 
%We therefore report, for the first time, also the presence of an outflow in the lower ionization [OII] emission and in the neutral component.

%°°°°Balmer decrements°°°°% 
%Balmer decrements were used to estimate the level of extinction in the two targets (see Section 5.2), which turned out to be fairly high for the components we associate to the outflows (A$_V$=1.8 for XID2028 and A$_V>0.5$ for XID5321; ``broad'' column in Table 2). This implies corrections of the order of $\sim6$ and $\sim2$ to the observed [OIII] line luminosities to estimate the intrinsic [OIII] line luminosities. This information is essential  to derive extinction corrected (broad) line luminosities and therefore a more precise estimate of the kinetic power associated to the ionised components.

The interpretation of the data we have for XID2028 is relatively simple.
The absorption lines of XID2028, tracing the low ionization/neutral gas, show a blueshift nearly like that observed in the emission lines, with $V_S\sim$ -300 km s$^{-1}$ with respect to the systemic velocity, easily explained as outflowing gas along the line of sight. 
%°°°°[OII]°°°°% 
 The [OII] emission doublet as sampled by the Keck/DEIMOS spectrum (Figure ~\ref{fitOIIK}) %were also fitted using the non-parametric approach.
% The comparison between the [OIII]$\lambda$5007 and [OII] doublets are shown in Figure ~\ref{profilivel}: [OII] velocity profiles are narrower than [OIII], but 
also shows almost the same asymmetries and shifts, within the errors, 
of [OIII] (Figure ~\ref{profilivel}; see also Zakamska \& Greene 2014 for similar [OII] and [OIII] broad profiles observed in SDSS spectra). 
% In XID2028, the [OII] doublet as sampled by the Keck/DEIMOS spectrum (Figure ~\ref{fitOIIK}) implies a velocity shift lower than the [OIII] velocity shift, but consistent within the errors.
With respect to the analysis presented in B14 and Cresci et al. (2014) we therefore report, for the first time in this prototypical system, also the presence of an outflow in the lower ionization [OII] emission and in the neutral component.

Under the assumptions discussed in Section \ref{secneutral}, we obtain a lower limit on the outflow mass rate for the neutral component that, added to the ionized component, gives us a lower limit on the outflow mass rate, $\dot M_{out}^{tot}(2028)>630\ M_{\odot}\ yr^{-1}$, without taking into account the presence of a possible molecular component. We also obtain a kinetic power associated to the outflow of $\sim$ $4 \cdot 10^{44}$ erg s$^{-1}$, where, starting from Equation \ref{canodiaz} for the ionized kinetic power, we have taken into account a correction factor of 10 from the comparison between the [OIII] and H$\beta$ mass rate estimates in the nuclear region %, and a further factor of 10 for the molecular component 
(see Section \ref{secioniz}). Although obtained through different observational constraints, these results are overall consistent with those presented in Cresci et al. (2014) and unambiguously point towards an AGN origin for the observed wind (see Section \ref{secioniz}).

\subsection{The redshifted outflow in XID5321}\label{sec5321outflow}

%The results regarding the [OII] and magnesium line, and the outflow properties of XID5321, showing a significantly diversity with respect to the XID2028 ones, will be discussed separately in the next Section.

For XID5321 the interpretation of the observational constraints we obtained in our analysis (in particular, the redshifted absorption that shows a velocity shift of the same magnitude of the emission lines) is more complex.
% An absorbing medium on the line of sight between the (ionising continuum) source and the gas responsible for the absorption line can reveal itself as redshifted outflow if it is rotationally dominated (Hall et al. 2013). Objects with redshifted and blueshifted absorptions are rare, but objects with only reshifted absorptions are really rare.

%In fact, XID5321 shows a great spatial extension, allowing to study the [OII] emission lines in the three apertures (Figure ~\ref{fitOIIX}). It turns out that the [OII] profiles tend to be wider and more redshifted from $a$ to $c$.

\begin{figure}
\centering
\includegraphics[scale=0.3]{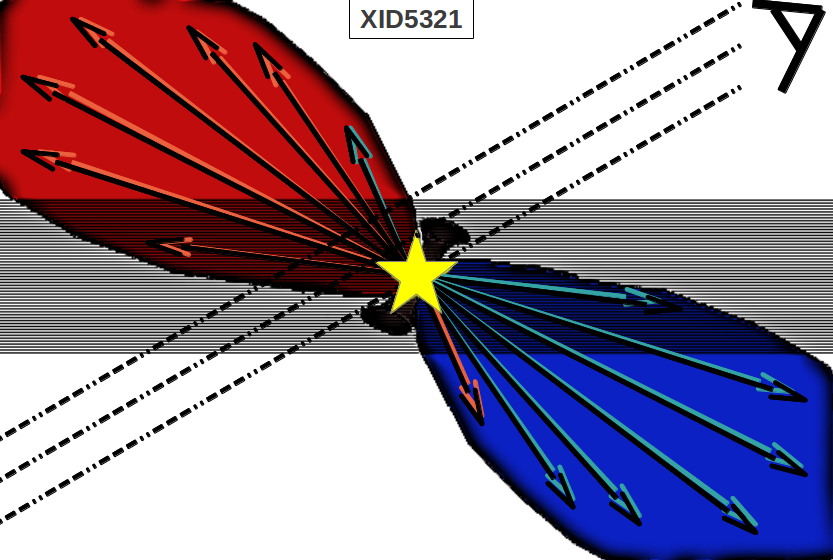}
%\captionsetup{font={small}}
\caption{Schematic cartoon showing the extended host galaxy (grey area) and conical outflow of XID5321 (blue and red areas). The blue and red lobes represents the blueshifted and redshifted gas. The yellow star and the cylindrical shape show the position of the driving center and the torus. The line of sight is indicated.}
\label{cartoon5321}
\end{figure}

The similarity between the velocity shift of absorption and ionized lines suggests that neutral and ionized gas are closely connected: in fact, some of the ionized outflowing gas could become again neutral \citep[e.g.,][]{Emonts2005}. %({\bf have a look to the figure in Bolatto et al. 2013, Nature.. there is a stratification of components..)}. 
In this case, the redshifted absorption lines observed in XID5321 could be explained as outflowing absorbers along the line of sight, illuminated by the light of the host galaxy behind the redshifted outflowing material. Figure \ref{cartoon5321} shows a cartoon illustrating the extended host galaxy disk (grey area) and our inferred geometry for the bi-conical outflow (blue and red areas) of XID5321. In this scenario, the observed outflow should corresponds to the receding red cone, while the blue cone is obscured by the host galaxy. In addition, the AGN system and the associated torus %, indicated with the central yellow star and the roughly cylindrical shape respectively, 
are warped with respect to the galaxy disk. 

%{\bf In the upper part of the figure, the positive velocity components prevails, while the negative components in predominant in the lower part of the figure ( true?? isn't it the opposite? why do not reromve this sentence..?)}.
An outflowing absorbing medium on the line of sight (indicated with dotted-dashed lines in the figure), in or close to the red cone and illuminated by the light of the extended host galaxy behind, could be responsible for the absorption lines observed, with the same kinematics and extension of the ionized outflowing gas.

The cartoon also helps us to understand the origin of the broad wings observed in the [OIII]$\lambda$5007 emission line (Figure \ref{aperture2D}). A blue wing could be originated by the near-side of the outflow (in the upper part of the figure), where the velocities have negative projected components along the line of sight. For this reason, these velocities are indicated with cyan arrows. The bulk has instead velocities with positive line of sight components (orange arrows) and determines the core of the line profile. An alternative option is that the bi-conical outflow has a more wider opening angle in such a way that the blue wing is originated by the blue cone extended up to the host galaxy.
 In these possible scenarios, the positive maximum velocity, $v_{max}$, is indicative of the outflow velocity because it is originated by the gas components at the periphery of the visible cone and closest to the line of sight. All the other gas components have lower velocities because of greater projection effects.

In addition, the proposed geometry also explains why we see the BLR responsible for the emission of the Broad H$\alpha$ (i.e., we have an almost unobscured view to the nucleus), why the broad H$\beta$ is heavily extincted (i.e., probably by the galaxy disk; see Table 2), and why we measure a relatively low X--ray column density (i.e., the line of sight does not fully intercept the torus), therefore justifying the type 1.8 of the source.  \\ 

XID5321 [OII] lines  show a great spatial extension, allowing us to study the emission in the three apertures (Figure ~\ref{fitOIIX}). It turns out that the [OII] profiles tend to be wider and more redshifted from $a$ to $c$. A possible explanation is that the low ionization component of the outflow could have a wider opening angle than the highly ionization component. 

%How to put the [OII] emission in this (or another) scenario?}\\

Following the same arguments mentioned above for XID2028, we obtain a kinetic power associated to the outflow of $\sim 6 \cdot 10^{44}$ erg s$^{-1}$. Following the scenario presented in Figure \ref{cartoon5321}, a lower limit on the outflow mass rate for the neutral component added to the ionized term was obtained: $\dot M_{out}^{tot}(5321)>535\ M_{\odot}\ yr^{-1}$.

Although the scenario described above seems to fit our diverse constraints and is favored, we should notice that the redshifted systems of XID5321 could in principle also be consistent with the existence of an object approaching the AGN, and/or with a kinematics disturbed by a double nucleus in the galaxy. This second scenario may be suggested by the double peak in the H$\alpha$ region (see Figure ~\ref{fitsimultanei}): the narrow peak we associate with the [NII]$\lambda$6583 could instead be the H$\alpha$ of a second source blended in the X-shooter spectrum (within 0.8'' and therefore within $\sim6$ kpc from the brightest AGN).
 In this case, we should see at least a second shifted [OIII]$\lambda$5007 emission line too, at observed wavelengths 1240.5 nm. A closer view of the spectrum presented in Figure ~\ref{aperture2D} does not allow us to exclude the presence of a weak emission line and high resolution imaging with HST will be vital to disentangle the nature of this source (see, e.g., \citealt{Greene2012}). %{\bf can we may use also the $w40$ information in the discussion of a possible double source?}

\subsection{Outlook}
The observed extents of the outflows for our two targets, although obtained through NIR integrated spectra, are among the largest reported in the literature for high-z luminous QSOs measured from IFU data. For example, \citet{CanoDiaz2012} reported clear evidence of outflowing gas on a much smaller scale ($\sim3$ kpc, on scale even lower than those sampled by our central apertures). Similarly, Alexander et al. (2010) detected broad and shifted components in J1237 out to 4-8 kpc within the galaxy, on scales comparable to our nuclear apertures. This highlights the power of high resolution slit spectroscopy in unambiguously revealing complex kinematics and assessing the order of magnitude of the energetics of the systems, when applied to luminous QSOs. Larger programs of NIR follow-up of large area and deep fields X--ray selected sources (such as those from XXL and Stripe82) can be exploited to single out the most promising cases.

Differently from our two targets, luminous QSOs detected in large area IR surveys seem to be ``X--ray weak'': the X--ray weakness can be ascribed to the high column density along the line of sight, as proposed for three WISE-selected QSOs at z$\sim2$ (Stern et al. 2014), or possibly to an intrinsically low disk/corona emission, as observed in J1234+0907(\citealt{Banerji2014}). In order to have a complete picture of the complex interplay between the X--ray luminosity, the X--ray absorption, and the winds properties, it is essential to gather a complete sample of sources with deep optical, NIR and X--ray data over a  large luminosity range.  A crucial role will be played by follow-up observations with PdBI and ALMA, aimed to detecting the molecular gas components expected to be associated to the outflows. As a first step, we obtained PdBI observations of the CO(3-2) transition in XID2028 with the aim of measuring the gas mass in the system.  

 % Harrison et al. (2012), analysing a larger sample (8 objects) of z$\sim2$ ULIRGs, probed also outflows on larger spatial scales, up to 20$ kpc. However, in their sources…

Finally, although the results of our study have shown that slit-resolved spectroscopy is more adequate than a single aperture nuclear spectra to put constraints on the properties of the outflows, we have found structural complexities for which integral field spectroscopy is required, especially in the case of XID5321 where possible accelerating outflow, or double nucleus could be present. Moreover, spatially resolved NIR spectroscopy can also be used to probe the impact of outflows on the host galaxy (see, e.g., \citealt{CanoDiaz2012}).

%observational proofs of the impact that the outflow observed in XID5321 could have on the host galaxy is still missing. In order to investigate this issue, we obtained SINFONI data (PI: Cresci) to map in seeing limited mode both H$\alpha$ and [OIII] regions in XID5321 and to study the dust extinction in combination with the already available J-band data in XID2028.
%for XID5321 we will test if the [OIII] line  possible non-detected component of the [OIII] lines could be present in the lateral $a$ aperture as well (?)).\\

\section*{Acknowledgments}

This work is based on observations made at the European Southern Observatory, Paranal, Chile (ESO program 090.A-0830(A)) and with Keck/DEIMOS, and on observations obtained with XMM-{\it Newton} and {\it Herschel}, two ESA Science Missions with instruments and contributions directly funded by ESA Member States and the USA (NASA).
MP and MB acknowledge support from the FP7 Career Integration Grant ``eEASy'' (``SMBH evolution through cosmic time: from current surveys to eROSITA-Euclid AGN Synergies'', CIG 321913), and the DFG cluster of excellence ``Origin and Structure of the Universe''.
Support for this publication was provided by the Italian National Institute for Astrophysics (INAF) through PRIN-INAF 2011 (``Black hole growth and AGN feedback through the cosmic time'') and by the Italian ministry for school, university and research (MIUR) through PRIN-MIUR 2010-2011 ``The dark Universe and the cosmic evolution of baryons: from current surveys to Euclid''. %and PRIN-INAF-2012 (``The life cycle of early black holes''), and by the Italian ministry for school, university and research (MIUR) through PRIN-MIUR 2010-2011 (``The dark Universe and the cosmic evolution of baryons: from current surveys to Euclid'').
We gratefully acknowledge the unique contribution of the entire COSMOS collaboration for making their excellent data products publicly available; more information on the COSMOS survey is available at \verb+http://www.astro.caltech.edu/~cosmos+.

%\appendix

%\newpage
%\bsp
\label{lastpage}
\end{document}